\documentclass[10pt]{article}

\usepackage{amsmath}
\usepackage{amsfonts}
\usepackage{amsmath,amssymb}
\usepackage{amssymb, amscd, amsthm}
\usepackage{graphicx}
\usepackage{color}
\usepackage{url}
\usepackage{longtable}
\usepackage{booktabs}
\usepackage{tabularx}
\renewcommand{\arraystretch}{1.3}

\newtheorem{theorem}{Theorem}
\newtheorem{example}{Example}
\newtheorem{lemma}{Lemma}
\newtheorem{corollary}{Corollary}
\newtheorem{proposition}{Proposition}
\newtheorem{remark}{Remark}
\usepackage[hidelinks]{hyperref}
\newcommand{\F}{\mathbb F}
\newcommand{\Tr}{\operatorname{Tr}}

\begin{document}
\title{Cross Correlation of Binary Niho-type Sequences and Related Constacyclic Codes}

\author{Yuehui Cui, Jinquan Luo, Yongbo Xia, Qiuyi Zhang}
\date{}
\maketitle
\insert\footins{\small{\it } Y. Cui, J. Luo, and Q, Zhang are with School of Mathematics and Statistics, Hubei Key Laboratory of Mathematical Sciences, Central China Normal University, Wuhan China 430079. (E-mail: hfcyh1@163.com; luojinquan@ccnu.edu.cn; 15750258308@163.com)

Y. Xia is with the Department of Mathematics and Statistics, South-Central Minzu University, Wuhan 430074, China. (E-mail: xia@mail.scuec.edu.cn)}
{\centering\section*{Abstract}}
 \addcontentsline{toc}{section}{\protect Abstract}
 \setcounter{equation}{0}

Let \(q=2^m\), where \(m\ge3\) is odd, and let \(d=(q-1)(q+4)/6+1\). We study the cross correlation between a binary \(m\)-sequence of length \(q^2-1\) and its \(d\)-decimated sequence of length \(3(q-1)\). We prove that the cross correlation is exactly four-valued and determine its distribution explicitly. For shift decimated sequence, we obtain at most five cross correlation values and express their frequencies in terms of a specified cubic root count over  \(\F_q\). We also determine the complete value distribution of the associated exponential sums. As an application, we construct \(q\)-ary constacyclic codes with parameters \([q+1,4,2(q+1)/3]_q\) and determine their weight distributions.

\medskip
\noindent{\large\bf Keywords: }\medskip Cross correlation; Niho exponent; exponential sum; constacyclic code; weight distribution.

\noindent{\bf2010 Mathematics Subject Classification}: 94A55, 11T23, 94B15.

\section{Introduction}\label{sec:introduction}

Sequences with low cross correlation are important in radar systems and CDMA communications. Background on sequence design and its applications can be found in \cite{H2,Golomb.Gong,niho,cdma,ROSEPHD}. Let \(s=(s_t)_{t=0}^{l-1}\) and \(s'=(s'_t)_{t=0}^{l-1}\) be two \(p\)-ary sequences of length \(l\), where \(p\) is prime. The cross correlation function of \(s\) and \(s'\)
at shift \(\tau\) is defined by
\begin{equation}\label{dfCss}
C_\tau(s,s')=\sum_{t=0}^{l-1}\zeta_p^{\,s_{t+\tau}-s'_t},\qquad 0\leq\tau<l,
\end{equation}
where \(\zeta_p=e^{2\pi\sqrt{-1}/p}\) is a primitive complex \(p\)-th root of unity, and sequence indices are interpreted modulo \(l\). The multiset \(\{C_\tau(s,s'):0\leq\tau<l\}\) is called the cross correlation distribution of \(s\) and \(s'\). In many cases, the sequence \(s'\) is a decimated sequence of \(s\) with some decimation \(d\), that is, \(s'_t=s_{dt\bmod l}\) for \(0\leq t<l\). An interesting problem in sequence design is to find decimations \(d\) that yield low cross correlation and to determine the possible values of \(C_\tau(s,s')\) and their frequencies. This problem has been studied extensively since the 1960s; see \cite{Canteaut.Charpin.Dobbertin,HellesethLi2024,Johansen.Helleseth.Kholosha,survey2,Zeng.Hu.Jiang.Yue.Cao} for known results, surveys, and related open problems.

If \(\gcd(d,l)=1\), then \(s'\) will have the same length as \(s\) and both these two sequences are called sequences of length \(l\). If \(\gcd(d,l)>1\), then  the length of \(s'\) will become \(l/\gcd(d,l)\) which is a factor of \(l\). In this case, cross correlation function of \(s\) and \(s'\) can still be defined  in \eqref{dfCss}.

Cross correlation between sequences of different lengths has also received considerable attention. Ness and Helleseth \cite{NessHelleseth2006a,NessHelleseth2006b} studied binary \(m\)-sequences of lengths \(2^{2m}-1\) and \(2^m-1\). For odd \(m\geq3\), they obtained three-valued cross correlation distributions using the decimations \((2^m+1)/3\) and \(2^{(m+1)/2}-1\), respectively. Helleseth, Kholosha, and Ness \cite{HellesethKholoshaNess2007} subsequently extended these results to a broader family and conjectured that their construction accounts for all three-valued cases for these two lengths. Luo \cite{L2} later constructed pairs of binary sequences of lengths \(2^{2m}-1\) and \(3(2^m-1)\) with three-valued cross correlation, where \(m\geq3\) is odd. For odd characteristic, Seo, Kim, No, and Shin \cite{4k} determined the cross correlation distributions between a \(p\)-ary \(m\)-sequence of length \(p^{4k}-1\) and its shift decimated sequences of length \(2(p^{2k}-1)\), with \(d=((p^{2k}+1)/2)^2\), where \(p\) is an odd prime and \(k\) is a positive integer. These studies demonstrate that sequences of unequal lengths provide another setting for obtaining few-valued cross correlation.

Niho-type decimations provide a useful algebraic setting for investigating these questions. A normalized Niho exponent \(d\) over \(\F_{p^{2m}}\) satisfies 
\[d\equiv1\pmod{p^m-1}.\]
The binary case was studied by Niho in his thesis \cite{niho}. Rosendahl \cite{Rosendahl2006} developed a generalization of Niho's theorem that provides a useful tool for determining the possible values of the associated exponential sums. In characteristic \(2\), the polar representation of nonzero elements of \(\F_{2^{2m}}\) reduces these sums to root-counting problems on the unit circle. Dobbertin, Felke, Helleseth, and Rosendahl \cite{DobbertinEtAl2006} used Dickson polynomials to study Niho-type cross correlation functions and established connections with Kloosterman sums. Table~\ref{table1} summarizes selected known results on few-valued cross correlation arising from Niho-type decimations.  Niho exponents have been extensively studied in connection with cross correlation, bent functions, linear codes, and permutation polynomials. For further details, see \cite{survey2,LiFengGe2014,LiXiongYan2026,xiong,Xie.Mesnager.Li.He.Zeng,xia1,xia2} and the references therein.

\begin{table}[!ht]
\centering
\caption{Known results on cross correlation arising from Niho-type decimations}
\label{table1}
\small
\renewcommand{\arraystretch}{1.35}
\setlength{\tabcolsep}{4pt}
\begin{tabularx}{\textwidth}{@{}l >{\raggedright\arraybackslash}X c c l@{}}
\toprule
\(d\) & Conditions
& \(\gcd(d,p^{n}-1)\) & Values & Reference\\
\midrule
\(2(p^m-1)+1\) & \(n=2m\), \(p^m\not\equiv2\pmod3\) & \(1\) & \(4\) & \cite{H2}\\
\(2(p^m-1)+1\) & \(n=2m\), \(p^m\equiv2\pmod3\)     & \(3\) & \(4\) & \cite{xias=2}\\
\(s(2^m-1)+1\) & \(p=2\), \(n=2m\),  \(4s\equiv1\pmod{2^m+1}\) & \(1\) & \(5\) & \cite{H2}\\
\(s(2^m-1)+1\) & \(p=2\), \(n=2m\), \(m\) odd, \(5s\equiv1\pmod{2^m+1}\) & \(3\) & \(4\) & \cite{L3}\\
\(s(2^m-1)+1\) & \(p=2\), \(n=2m\),  \(m/\gcd(m,r)\) even,\newline  \(s\equiv2^r(2^r\pm1)^{-1}\pmod{2^m+1}\) & \(1\) & \(4\) & \cite{DobbertinEtAl2006}\\
\(3(2^m-1)+1\) & \(p=2\), \(n=2m\), \(\gcd(5,2^m+1)=1\) & \(1\) & \(5\) or \(6\) & \cite{xia1}\\
\(3(3^m-1)+1\) & \(p=3\), \(n=2m\), \(m\not\equiv2\pmod4\) & \(1\) & \(5\) & \cite{xia2}\\
\(3(p^m-1)+1\) & \(p\geq5\), \(n=2m\), \(\gcd(5,p^m+1)=1\) & \(1\) & \(6\) & \cite{xiong}\\
\(4(2^m-1)+1\) & \(p=2\), \(n=2m\) & \(1\) & \(5\) or \(6\) & \cite{Xiong.Yan.niho4}\\
$\dfrac{p^m+1}{3}(p^m-1)+p^i$ & \(p\equiv2\pmod3\), \(n=2m\), \(0\leq i<2m\),\newline  \((p^{2m}-1)/(3p^i)\not\equiv2\pmod3\) & \(1\) & \(6\) & \cite{H2}\\
\(\dfrac{p^m+3}{4}(p^m-1)+1\) & \(p\) odd, \(n=2m\), \(p^m\equiv1\pmod4\) & \(\dfrac{p^m+1}{2}\) & \(4\) & \cite{L1}\\
\(\dfrac{2^m+4}{6}(2^m-1)+1\) & \(p=2\), \(n=2m\), \(m\geq3\) odd & \(\dfrac{2^m+1}{3}\) & \(4\) or \(5\) & This paper\\
\bottomrule
\end{tabularx}
\end{table}

To compare different decimations, it is convenient to express cross correlation functions in terms of exponential sums. Let \(\psi\) be a primitive element of \(\F_{p^n}\), and define
\begin{equation}\label{eq:intro-Sd}
S_d(a,b)=\sum_{x\in\F_{p^n}}\zeta_p^{\,\Tr_1^n(ax+bx^d)}, \qquad a,b\in\F_{p^n},
\end{equation}
where \(\Tr_1^n\) denotes the absolute trace from \(\F_{p^n}\) to \(\F_p\). For the \(m\)-sequence \(s_t=\Tr_1^n(\psi^t)\) and its \(d\)-decimated sequence, the substitution \(x=\psi^t\) gives \(C_\tau(s,s')=S_d(\psi^\tau,-1)-1\). Thus their cross correlation distribution is determined by the values of \(S_d(a,-1)\) as \(a\) ranges over \(\F_{p^n}^*\).

Two decimations \(d\) and \(e\) over \(\F_{p^n}\) are called equivalent if \(e\equiv p^j d\pmod{p^n-1}\) or \(e\equiv p^j d^{-1}\pmod{p^n-1}\) for some \(0\leq j<n\), where the latter requires \(\gcd(d,p^n-1)=1\). Trace invariance and a change of variables give \(S_{p^jd}(a,b)=S_d(a,b^{p^{n-j}})\) and \(S_{d^{-1}}(a,b)=S_d(b,a)\). These equivalences preserve the value distribution of \(S_d(a,b)\) over \(\F_{p^n}^2\), as well as the corresponding cross correlation distribution.

For example, if \(4s\equiv1\pmod{2^m+1}\) and \(e=s(2^m-1)+1\), then \(4e-(2^m+3)\equiv0\pmod{2^{2m}-1}\). Hence \(2^m+3\) is equivalent to the Niho-type decimation \(e\). Similarly, Mesnager and Wu \cite{MesnagerWu2025} determined the Walsh spectrum of \(x^{2\cdot3^m+1}\) over \(\F_{3^{2m}}\). We have \(2\cdot3^m-1\equiv3(2\cdot3^m+1)^{-1}\pmod{3^{2m}-1}\). Thus \(2\cdot3^m+1\) is equivalent to Niho-type decimation \(2(3^m-1)+1\).

Exponential sums also connect sequence correlation with coding theory through trace representations of codewords. In such constructions, the Hamming weight of a codeword can be expressed in terms of character sums, allowing correlation and weight distributions to be studied within a common framework.

Constacyclic codes provide a natural extension of cyclic codes. For a prime power \(q\) and an element \(\lambda\in\F_q^*\), an \(\F_q\)-linear code \(\mathcal C\subseteq\F_q^N\) is called \(\lambda\)-constacyclic if 
\[
(c_0,c_1,\ldots,c_{N-1})\in\mathcal C \quad\Longrightarrow\quad (\lambda c_{N-1},c_0,\ldots,c_{N-2})\in\mathcal C.
\]
The case \(\lambda=1\) gives a cyclic code. The Hamming weight distribution records the number of codewords of each weight and, in particular, determines the minimum distance. For codes given by trace representations, its determination often reduces to evaluating the associated exponential sums.

This connection has been used in several studies of codes with Niho-type exponents. Li, Feng, and Ge~\cite{LiFengGe2014} determined the weight distributions of three classes of cyclic codes with Niho exponents by evaluating associated exponential sums. Xiong, Li, Zhou, and Ding~\cite{XiongEtAl2016} studied two families with an arbitrary number of generalized Niho-type zeroes and determined their weight distributions. More recently, Li, Xiong, and Yan~\cite{LiXiongYan2026} determined the weight distributions of binary cyclic codes whose generalized Niho-type nonzeroes are indexed by exponents obtained from an arithmetic progression by omitting one term. Their approach combines polar representations, counting arguments involving binary Zetterberg codes, and power moments. These results motivate the study of specific Niho exponents whose root structure permits explicit computations in both sequence correlation and coding theory.

In this paper, let \(m\geq3\) be odd and \(q=2^m\). We consider the Niho exponent \(d=\frac{q+4}{6}(q-1)+1\) and write \(S(a,b)=S_d(a,b)\) for this exponent. We determine the four-valued cross correlation distribution of a binary \(m\)-sequence of length \(q^2-1\) and its \(d\)-decimated sequence. For shift decimated sequence, the five candidate frequencies are given in terms of \(\rho_c\) in Theorem~\ref{thm:shift-distribution}. This parameter counts the nonzero \(w\in\F_q\) for which exactly one of three specified cubics has three roots and the other two have none. When \(m=3k\) with \(k\) odd, we prove \(\rho_{(q+1)/9}=\rho_{2(q+1)/9}=(q+1-3\kappa_k)/18\), where \(\kappa_k\) is determined by the recurrence in Theorem~\ref{thm:rho-special-shifts}. The proof relates the cubic root counts to point counts on two curves over \(\F_8\).We then determine the complete value distribution of \(S(a,b)\) as \((a,b)\) ranges over \(\F_{q^2}^2\).

Using a relative-trace construction, we obtain a family of \(\lambda\)-constacyclic codes over \(\F_q\), where \(\lambda=\psi^{q+1}\) and \(\psi\) is a primitive element of \(\F_{q^2}\), with parameters \(\left[q+1,\ 4,\ \frac{2(q+1)}3\right]_q.\)  Their Hamming weight distributions are determined. For \(m\geq5\), these codes have six nonzero weights; for \(m=3\), the construction yields an \([9,4,6]_8\) maximum distance separable (MDS) code. Since \(\lambda\) has order \(q-1>2\), the codes are also Euclidean linear complementary dual (LCD), that is, \(\mathcal C\cap\mathcal C^\perp=\{\mathbf0\}\), by \cite[Corollary~3.3]{LiuFanLiu2018}.

The remainder of this paper is organized as follows. Section~\ref{sec:preliminaries} introduces the notation and preliminary results. Section~\ref{sec:correlation} determines the four-valued cross correlation distribution and studies shift decimated sequence. Section~\ref{sec:exponential-codes} establishes the value distribution of \(S(a,b)\) and presents the related constacyclic codes, including their weight distributions and LCD property. Section~\ref{conclusion} concludes the paper. 

\section{Preliminaries}\label{sec:preliminaries}
We fix the following notation for use from this section onward, unless otherwise stated.

\begin{tabular}{@{}c l l@{}}
\textbullet & \(q\) & \(q=2^m\), $m\geq3$ is an odd integer \\
\textbullet & \(h\), \(d\) & \(h=\frac{q+1}{3}\) and \(d=(q-1)\frac{q+4}{6}+1\)\\
\textbullet & \(\mathbb{F}_{q^2}\) & finite field of \(q^2\) elements\\
\textbullet & \(\mathbb{F}_{q^2}^*\) & multiplicative group of nonzero elements of \(\mathbb{F}_{q^2}\) \\
\textbullet & \(\psi\), \(\beta\) & \(\psi\) is a primitive element of \(\mathbb{F}_{q^2}\), and \(\beta=\psi^{(q^2-1)/3}\) \\
\textbullet & \(\bar x\) & \(\bar x=x^q\)\\
\textbullet & \(\Tr_{l_1}^{l_2}\) &  trace function from \(\F_{2^{l_2}}\) to \(\F_{2^{l_1}}\) for positive integers $l_1\mid l_2$\\
\textbullet & \(\chi(t)\) & \(\chi(t)=(-1)^{\Tr_1^m(t)}\)\\
\textbullet & \(\varepsilon\) &  \(\varepsilon=1\) if \(3\mid m\), and \(\varepsilon=0\) otherwise \\
\textbullet & \(U\), \(U_j\) & $U=\{u\in\F_{q^2}^*:u^{q+1}=1\}$, $U_j=\{u\in U:u^h=\beta^j\},\, j=0,1,2$\\
\textbullet & \(D\) & $D=\{x^d:x\in\F_{q^2}^*\}$\\
\textbullet & \(S(a,b)\) &  $S(a,b)=\sum_{x\in\F_{q^2}}(-1)^{\Tr_1^{2m}(ax+bx^d)}$, $a,b\in\F_{q^2}$\\
\textbullet & \(\#X\) & cardinality of a finite set \(X\)\\
\textbullet & \(R(w)\) & $R(w)=\#\{x\in\F_q:x^3+x=w\}$, \(w\in\F_q^*\)\\
\end{tabular}
\smallskip

We first record some elementary properties of the unit circle. Since \(m\) is odd, \(\beta^q=\beta^{-1}=\beta^2\) and \(\beta+\beta^2=1\). The group \(U\) is cyclic of order \(q+1=3h\). The homomorphism \(u\mapsto u^h\) maps \(U\) onto \(\{1,\beta,\beta^2\}\), with kernel of order \(h\). Consequently,
\[
U=U_0\cup U_1\cup U_2, \qquad \#U_j=h\quad(j=0,1,2).
\]
In particular, \(U_0\) is the subgroup of cubes in \(U\).

Since \(\F_q^*\) and \(U\) have relatively prime orders \(q-1\) and \(q+1\), their intersection is \(\{1\}\). Their product therefore has \((q-1)(q+1)=q^2-1\) elements and equals \(\F_{q^2}^*\). Thus every nonzero element has a unique polar representation
\begin{equation}\label{eq:polar}
x=ru,\qquad r\in\F_q^*,\quad u\in U.
\end{equation}

\begin{lemma}\label{lem:exponent}
The exponent \(d\) satisfies \(d\equiv1\pmod{q-1}\), \(d\equiv-h\pmod{q+1}\), and \(\gcd(d,q^2-1)=h\). Moreover,
\begin{equation}\label{eq:D}
D=\{x^d:x\in\F_{q^2}^*\}  =\langle\psi^h\rangle =\F_q^*\cup\beta\F_q^*\cup\beta^2\F_q^*, \qquad \#D=3(q-1).
\end{equation}
\end{lemma}

\begin{proof}
The congruence \(d\equiv1\pmod{q-1}\) follows immediately from the definition of \(d\). Since \(d=h\frac{q+2}{2}\), \(\frac{q+2}{2}\equiv2\pmod3\), and \(q+1=3h\), we obtain \(d\equiv-h\pmod{q+1}.\) Consequently, \(\gcd(d,q-1)=1\) and  \(\gcd(d,q+1)=h\gcd\left(\frac{q+2}{2},3\right)=h.\) As \(q-1\) and \(q+1\) are relatively prime, it follows that \(\gcd(d,q^2-1)=h\).

Since \(\F_{q^2}^*=\langle\psi\rangle\) is cyclic,
\[
D=\langle\psi^d\rangle=\langle\psi^{\gcd(d,q^2-1)}\rangle=\langle\psi^h\rangle.
\]
For \(x=ru\) as in \eqref{eq:polar}, the two congruences give \(x^d=ru^{-h}\). As \(r\) and \(u\) vary independently, \(r\) runs through \(\F_q^*\), while \(u^{-h}\) runs through \(\{1,\beta,\beta^2\}\). These three cosets of \(\F_q^*\) are distinct because \(3\nmid(q-1)\). This proves \eqref{eq:D}.
\end{proof}

We shall also use the identity
\begin{equation}\label{eq:S-scaling}
S(a,bv^d)=S(a/v,b), \qquad a,b\in\F_{q^2},\quad v\in\F_{q^2}^*,
\end{equation}
which follows from the substitution \(y=vx\). In particular, the value distribution of \(S(a,b)\), with \(a\) ranging over either \(\F_{q^2}\) or \(\F_{q^2}^*\), depends on a nonzero coefficient \(b\) only through the multiplicative coset \(bD\). Indeed, if \(b'=bv^d\), then \(S(a,b')=S(a/v,b)\), and multiplication by \(v^{-1}\) permutes both parameter sets. Since \(D=\langle\psi^h\rangle\), its index in \(\F_{q^2}^*\) is \(h\), and the distinct cosets are \(\psi^cD\), \(0\le c<h\).

We recall the following criterion for a quadratic equation to have both roots in the unit circle, together with a generalized Niho reduction that will be used in the subsequent sections.

\begin{lemma}[\cite{Tu}]\label{lem:unit-quadratic}
Let \(A,B\in\F_{q^2}^*\) satisfy \(\Tr_1^{2m}(B/A^2)=0\). Then both roots of \(X^2+AX+B=0\) belong to \(U\) if and only if
\begin{equation}\label{eq:unit-quadratic-condition}
B=\frac{A}{\bar A}\quad\text{and}\quad \Tr_1^m\left(\frac{B}{A^2}\right)=1.
\end{equation}
Under the first condition, \(B/A^2=1/(A\bar A)\in\F_q\), so the trace in the second condition is well defined.
\end{lemma}

\begin{lemma}[\cite{Rosendahl2006}]\label{lem:niho}
Let \(d_i=s_i(q-1)+1\), where \(s_i\) are nonnegative integers for \(i=1,2\). For fixed \(a,b\in\F_{q^2}\), let \(N\) denote the number of solutions \(z\in U\) of
\begin{equation}\label{eq:general-niho-equation}
\bar b z^{2s_2-1}+\bar a z^{s_1+s_2-1}+a z^{s_2-s_1}+b=0.
\end{equation}
Then
\[
\sum_{x\in\F_{q^2}^*}(-1)^{\Tr_1^{2m}(ax^{d_1}+bx^{d_2})}=-1+q(N-1).
\]
\end{lemma}

For our particular exponent, the preceding lemma takes the following form, which will be used throughout the paper.

\begin{lemma}\label{lem:unit-reduction}
For \(b\in\F_{q^2}\), put
\begin{equation}\label{eq:shift-tj}
t_j=t_j(b)=\bar b\beta^j+b\beta^{-j}\in\F_q,\qquad j=0,1,2.
\end{equation}
For \(a,b\in\F_{q^2}\), define
\begin{equation}\label{eq:unit-root-count}
V(a,b)=\sum_{j=0}^{2}\#\{z\in U_j:az^2+t_jz+\bar a=0\}.
\end{equation}
Then
\begin{equation}\label{eq:shift-S-V}
S(a,b)=q\bigl(V(a,b)-1\bigr).
\end{equation}
Moreover,
\begin{equation}\label{eq:t-properties}
t_0+t_1+t_2=0, \qquad t_j=0\ \Longleftrightarrow\ b\beta^{-j}\in\F_q.
\end{equation}
Thus, for \(b\in\F_{q^2}^*\setminus D\), the three coefficients \(t_0,t_1,t_2\) are nonzero and pairwise distinct.
\end{lemma}

\begin{proof}
Apply Lemma~\ref{lem:niho} with \(s_1=0\) and \(s_2=(q+4)/6\). Squaring permutes \(U\), whose order \(q+1\) is odd. Replacing the variable in \eqref{eq:general-niho-equation} by \(z^2\), and observing that \(2(2s_2-1)=2h\), \(2(s_2-1)=h-1\), and \(2s_2=h+1\), we obtain the equation
\begin{equation}\label{eq:special-niho-equation}
\bar b z^{2h}+\bar a z^{h-1}+az^{h+1}+b=0, \qquad z\in U.
\end{equation}
If \(z\in U_j\), multiplication by \(z\beta^{-j}\) turns this equation into
\begin{equation}\label{eq:shift-quadratic}
az^2+t_jz+\bar a=0.
\end{equation}
The three cosets partition \(U\), so their root counts add to \(V(a,b)\). Adding the term \(x=0\) to the sum in Lemma~\ref{lem:niho} proves \eqref{eq:shift-S-V}.

Since \(\beta^q=\beta^{-1}\), we have \(t_j=(b\beta^{-j})^q+b\beta^{-j}\in\F_q\). This relative trace is zero precisely when \(b\beta^{-j}\in\F_q\), and summing over \(j\) gives \(t_0+t_1+t_2=0\). For \(b\ne0\), a zero \(t_j\) is therefore equivalent to \(b\in\beta^j\F_q^*\). For \(b\in\F_{q^2}^*\setminus D\), none of the \(t_j\) vanishes. Since their sum is zero, equality of any two would force the third to vanish. They are therefore pairwise distinct.
\end{proof}

We recall a special case of Bluher's theorem on polynomial root counts, which will be used to derive the frequency formulas in Lemma~\ref{lem:cubic}.

\begin{lemma}[{\cite[Theorem~5.6]{Bluher2004}}]\label{lem:bluher-cubic}
Let \(q=2^m\), where \(m\) is odd. For every \(b\in\F_q^*\), the polynomial \(X^3+bX+b\) has either \(0\), \(1\), or \(3\) roots in \(\F_q\). Moreover,
\[
\begin{aligned}
&\#\bigl\{b\in\F_q^*:  X^3+bX+b \text{ has no root in }\F_q\bigr\}=\frac{q+1}{3},\\
&\#\bigl\{b\in\F_q^*:  X^3+bX+b \text{ has exactly one root in }\F_q\bigr\}=\frac{q-2}{2},\\
&\#\bigl\{b\in\F_q^*: X^3+bX+b \text{ has exactly three roots in }\F_q\bigr\}=\frac{q-2}{6}.
\end{aligned}
\]
\end{lemma}

Next, we recall a basic fact, which will be useful for the subsequent proof. For any \(u\in\F_{2^n}\), the number of solutions to \(y^2+y=u\) is given by
\begin{equation}\label{eq:trace-fibres}
 \#\{y\in\F_{2^n}:y^2+y=u\}= \begin{cases}2,&\Tr_1^n(u)=0,\\0,&\Tr_1^n(u)=1.\end{cases} 
\end{equation}

The following lemma relates \(R(w)\) to the partition \(U=U_0\cup U_1 \cup U_2\).

\begin{lemma}\label{lem:cubic}
For every \(w\in\F_q^*\), we have \(R(w)=\#\{x\in\F_q:x^3+x=w\}\in\{0,1,3\}\). More precisely, the following classification holds:
\begin{center}
\renewcommand{\arraystretch}{1.2}
\begin{tabular}{@{}clc@{}}
\toprule
\(R(w)\) & Roots of \(Z^2+wZ+1\) & \(\chi(1/w)\)\\
\midrule
\(1\) & Both in \(\F_q^*\), neither in \(U\) & \(1\)\\
\(3\) & Both in \(U_0\) & \(-1\)\\
\(0\) & One in \(U_1\) and one in \(U_2\) & \(-1\)\\
\bottomrule
\end{tabular}
\end{center}
Furthermore,
\begin{equation}\label{eq:cubic-distribution}
\#\{w\in\F_q^*:R(w)=1\}=\frac{q-2}{2},\,\,
\#\{w\in\F_q^*:R(w)=3\}=\frac{q-2}{6},\,\,
\#\{w\in\F_q^*:R(w)=0\}=\frac{q+1}{3}.
\end{equation}
\end{lemma}
\begin{proof}
For any \(t\in\F_q^*\), we have \(\Tr_1^{2m}(1/t^2)=0\) and \(t/\bar t=1\). Thus Lemma~\ref{lem:unit-quadratic}, applied with \(A=t\) and \(B=1\), shows that the two roots of \(Z^2+tZ+1\) belong to \(U\) if and only if \(\Tr_1^m(1/t)=1\). Here we use \(\Tr_1^m(1/t^2)=\Tr_1^m(1/t)\). If this trace is zero, substituting \(Z=tT\) gives \(T^2+T=t^{-2}\). Equation~\eqref{eq:trace-fibres} then gives two roots in \(\F_q\), both nonzero. Neither then belongs to \(U\), since \(\F_q^*\cap U=\{1\}\) and \(1\) is not a root. In either case, the roots are distinct and reciprocal.

We next relate the roots of \(Z^2+wZ+1\) to those of \(X^3+X+w\) using the identity \begin{equation}\label{eq:dickson-identity} 
(v+v^{-1})^3+(v+v^{-1})=v^3+v^{-3}.
\end{equation}
Every solution \(x\in\F_q\) of \(x^3+x=w\) is nonzero. Applying the preceding quadratic-root classification with \(t=x\), we can write \(x=v+v^{-1}\) for some \(v\in\F_q^*\cup U\). By \eqref{eq:dickson-identity}, \(v^3\) is a root of \(Z^2+wZ+1\). Conversely, every such \(v\) gives a solution \(x=v+v^{-1}\in\F_q\) of \(x^3+x=w\). 

Two admissible elements \(v,v'\) give the same \(x\) if and only if \(v'=v\) or \(v'=v^{-1}\), since they satisfy the same quadratic \(V^2+xV+1=0\). Moreover, \(v=v^{-1}\) would imply \(v=1\) and hence \(w=0\), a contradiction. Therefore, \(R(w)\) equals half the number of admissible elements \(v\).

If the roots of \(Z^2+wZ+1\) belong to \(\F_q^*\), each has exactly one cube root in \(\F_q^*\), because \(\gcd(3,q-1)=1\). Neither has a cube root in \(U\), so there are exactly two admissible elements and \(R(w)=1\). The trace criterion above gives \(\chi(1/w)=1\).

If the quadratic roots belong to \(U\), neither has a cube root in \(\F_q^*\), since neither root is \(1\). The cubing map on \(U\) has kernel \(\{1,\beta,\beta^2\}\) and image \(U_0\). Consequently, if both roots belong to \(U_0\), each has three cube roots in \(U\), giving \(R(w)=3\). Otherwise, their reciprocal relation places one in \(U_1\) and the other in \(U_2\); neither has a cube root in \(U\), giving \(R(w)=0\). In both cases, \(\chi(1/w)=-1\). This proves the classification.

Finally, the substitution \(x=wy\) gives
\[
x^3+x+w=w^3\bigl(y^3+w^{-2}y+w^{-2}\bigr).
\]
Thus \(R(w)\) equals the number of roots of \(Y^3+bY+b\) in \(\F_q\), with \(b=w^{-2}\). Since \(w\mapsto w^{-2}\) permutes \(\F_q^*\), the frequencies in \eqref{eq:cubic-distribution} follow from Lemma~\ref{lem:bluher-cubic}.
\end{proof}

\section{Cross correlation distribution}\label{sec:correlation}
In this section, we determine the cross correlation distribution for the \(d\)-decimated sequence and then study the corresponding shift decimated sequence.

\subsection{Four-valued cross correlation}\label{subsec:four-valued}
In this subsection, we will study cross correlation of sequences 
\begin{equation}\label{s}
  s=(s_t)_{t=0}^{q^2-2}=(\Tr_1^{2m}(\psi^t))_{t=0}^{q^2-2}
\end{equation}
and its \(d\)-decimated sequence
\begin{equation}\label{s'}
  s'=(s'_t)_{t=0}^{q^2-2}=(s_{dt})_{t=0}^{q^2-2}=(\Tr_1^{2m}(\psi^{dt}))_{t=0}^{q^2-2}
\end{equation}
with decimation $d$. Then by (\ref{dfCss}), the cross correlation function with shift $\tau$ is
\begin{equation}\label{eq:cross-S}
C_\tau(s,s') =\sum_{t=0}^{q^2-2}(-1)^{s_{t+\tau}+s'_t}=\sum_{x\in\F_{q^2}^*} (-1)^{\Tr_1^{2m}(ax+x^d)}=S(a,1)-1,
\qquad a=\psi^\tau.
\end{equation}
As \(\tau\) ranges from \(0\) to \(q^2-2\), \(a\) runs through
\(\F_{q^2}^*\) exactly once. We first determine the possible cross correlation values.
\begin{lemma}\label{lem:four-values}
For every \(a\in\F_{q^2}^*\),
\[
C_\tau(s,s')\in\{-1-q,-1,-1+q,-1+2q\}, \qquad a=\psi^\tau.
\]
In addition,
\begin{equation}\label{eq:S01}
S(0,1)=q(h-1)=\frac{q(q-2)}3.
\end{equation}
\end{lemma}

\begin{proof}
For \(b=1\), the coefficients in \eqref{eq:shift-tj} are \(t_0=0\) and \(t_1=t_2=1\). Consequently, Lemma~\ref{lem:unit-reduction} gives
\begin{equation}\label{eq:b1-root-count}
\begin{aligned}
V(a,1) ={}\#\{z\in U_0:az^2+\bar a=0\}+\#\{z\in U_1\cup U_2:az^2+z+\bar a=0\}.
\end{aligned}
\end{equation}
When \(a\ne0\), the first equation has exactly one root in \(\F_{q^2}\), because squaring is a permutation of this field. The second equation has at most two roots. Thus \(0\le V(a,1)\le3\), and the possible cross correlation values follow from \eqref{eq:shift-S-V} and \eqref{eq:cross-S}.

When \(a=0\), every element of \(U_0\) satisfies the first equation, while the second equation has no root in \(U\). Hence \(V(0,1)=h\), proving \eqref{eq:S01}.
\end{proof}

We next determine the frequency of the value \(-1-q\).
\begin{lemma}\label{lem:Nminus}
We have
\begin{equation}\label{eq:Nminus}
\#\{a\in\F_{q^2}^*:S(a,1)=-q\} =\frac{(q+1)(q-2)}3.
\end{equation}
\end{lemma}

\begin{proof}
By \eqref{eq:shift-S-V}, it suffices to count the nonzero \(a\) for which \(V(a,1)=0\). Write \(a=ru\), with \(r\in\F_q^*\) and \(u\in U\). Since \(\bar a=ru^{-1}\), the unique root of \(az^2+\bar a=0\) is \(z=u^{-1}\). It belongs to \(U_0\) exactly when \(u\in U_0\). Under the substitution \(y=uz\), the second equation in \eqref{eq:b1-root-count} becomes
\begin{equation}\label{eq:b1-transformed}
y^2+\frac1r\,y+1=0.
\end{equation}

If \(R(1/r)=1\), Lemma~\ref{lem:cubic} shows that \eqref{eq:b1-transformed} has no root in \(U\). Since multiplication by \(u\) permutes \(U\), the second term in \eqref{eq:b1-root-count} is zero. Therefore \(V(a,1)=0\) exactly when \(u\notin U_0\). There are \((q-2)/2\) choices of \(r\) of this type and \(2h\) choices of \(u\notin U_0\).

It remains to rule out \(V(a,1)=0\) in the other two cases. If \(R(1/r)=3\), both roots \(y\) belong to \(U_0\). When \(u\in U_0\), the first term of \eqref{eq:b1-root-count} is already one; otherwise, both roots \(z=y/u\) belong to \(U_1\cup U_2\), so the second term is two. If \(R(1/r)=0\), one root \(y\) belongs to \(U_1\) and the other to \(U_2\). These roots cannot both lie in the single coset \(uU_0\). Hence at least one corresponding \(z=y/u\) lies in \(U_1\cup U_2\), again giving a positive contribution.

Thus the required number is \(2h(q-2)/2=(q+1)(q-2)/3\).
\end{proof}

The remaining frequencies will follow from the next two moment identities, for which we include a short proof.
\begin{lemma}\label{lem:fixed-b-moments}
For every fixed \(b\in\F_{q^2}\),
\begin{equation}\label{eq:fixed-b-moments}
\sum_{a\in\F_{q^2}}S(a,b)=q^2, \qquad \sum_{a\in\F_{q^2}}S(a,b)^2=q^4.
\end{equation}
\end{lemma}

\begin{proof}
Additive-character orthogonality gives
\[
\sum_{a\in\F_{q^2}}(-1)^{\Tr_1^{2m}(at)}=
\begin{cases}
q^2,&t=0,\\
0,&t\ne0.
\end{cases}
\]
Interchanging the sums in the first moment, only \(x=0\) contributes, giving \(\sum_{a\in\F_{q^2}}S(a,b)=q^2\).

For the second moment, expanding the square introduces two independent summation variables \(x,y\in\F_{q^2}\):
\[
S(a,b)^2 =\sum_{x,y\in\F_{q^2}} (-1)^{\Tr_1^{2m}(a(x+y)+b(x^d+y^d))}.
\]
Consequently,
\[
\sum_{a\in\F_{q^2}}S(a,b)^2=\sum_{x,y\in\F_{q^2}}(-1)^{\Tr_1^{2m}(b(x^d+y^d))}\sum_{a\in\F_{q^2}}(-1)^{\Tr_1^{2m}(a(x+y))}.
\]
By orthogonality, the inner sum vanishes unless \(x+y=0\), which is equivalent to \(x=y\) in characteristic \(2\). For each of the \(q^2\) pairs \((x,x)\), we have \(b(x^d+x^d)=0\), and the inner sum equals \(q^2\). Thus the second moment is \(q^4\).
\end{proof}

We can now determine the complete cross correlation distribution.
\begin{theorem}\label{thm:cross-distribution}
As \(\tau\) ranges from \(0\) to \(q^2-2\), the cross correlation distribution of \(s\) and \(s'\) is
\[
C_\tau(s,s')=
\left\{
\begin{array}{lll}
-1,     & \text{occurs }2(2q^2+q-1)/9   &\text{times},\\
-1-q,   & \text{occurs }(q^2-q-2)/3     &\text{times},\\
-1+q,   & \text{occurs }(q^2-q-2)/9     &\text{times},\\
-1+2q,  & \text{occurs }(q+1)^2/9       &\text{times}.
\end{array}
\right.
\]
\end{theorem}
\begin{proof}
For \(i=-1,0,1,2\), let \(M_i=\#\{a\in\F_{q^2}^*:S(a,1)=iq\}\). By \eqref{eq:S01}, \eqref{eq:Nminus} and Lemma~\ref{lem:fixed-b-moments}, we obtain
\begin{equation}\label{eq:thm1-4}
\begin{aligned}
&  M_{-1} =(q+1)(q-2)/3,\\
&  M_{-1}+M_0+M_1+M_2=q^2-1,\\
&  -M_{-1}+M_1+2M_2=q-(h-1),\\
&  M_{-1}+M_1+4M_2=q^2-(h-1)^2.
\end{aligned}
\end{equation}
Solving this system and using \(C_\tau(s,s')=S(\psi^\tau,1)-1\) gives the stated distribution.
\end{proof}

\subsection{On shift decimated sequence}\label{subsec:shifted-decimation}
We next consider the shift decimated sequence
\[
s'_c=(s_{dt+c})_{t=0}^{q^2-2} =\left(\Tr_1^{2m}(\psi^{dt+c})\right)_{t=0}^{q^2-2},\qquad 0\le c<h.
\]
Its cross correlation with \(s\) is
\begin{equation}\label{eq:shift-cross-S}
C_\tau(s,s'_c)=\sum_{t=0}^{q^2-2}(-1)^{\Tr_1^{2m}(\psi^{t+\tau}+\psi^{dt+c})}=\sum_{x\in\F_{q^2}^*}
(-1)^{\Tr_1^{2m}(ax+bx^d)}=S(a,b)-1,
\end{equation}
where \(a=\psi^\tau\) and \(b=\psi^c\). The case \(c=0\) has been settled in Theorem~\ref{thm:cross-distribution}. For \(1\le c<h\), we have \(b\notin D\), since \(D=\langle\psi^h\rangle\).  We again begin by determining the possible cross correlation values.
\begin{lemma}\label{lem:shift-values}
For every \(0\le c<h\) and every \(\tau\),
\[
C_\tau(s,s'_c)\in\{-1-q,-1,-1+q,-1+2q,-1+3q\}.
\]
Moreover, \(S(0,\psi^c)=-q\) whenever \(1\le c<h\).
\end{lemma}
\begin{proof}
The case \(c=0\) follows from Lemma~\ref{lem:four-values}. More generally, consider any \(b\in\F_{q^2}^*\setminus D\). By Lemma~\ref{lem:unit-reduction}, all three \(t_j\) are nonzero and \(V(a,b)\) is given by \eqref{eq:unit-root-count}.

For \(a\ne0\), each quadratic contributes at most two roots in its prescribed coset. Moreover, two distinct cosets cannot both contribute two roots. Indeed, if the quadratic for \(U_j\) has two roots in \(U_j\), their product is \(\bar a/a\), by Vieta's formulas, and hence \((\bar a/a)^h=\beta^{2j}.\) The same conclusion for a distinct index \(k\) would give \(\beta^{2j}=\beta^{2k}\), which is impossible. Thus at most one coset contributes two roots and \(V(a,b)\le2+1+1=4\). Equation~\eqref{eq:shift-S-V} now gives the asserted set of possible values.

For \(a=0\), the equations become \(t_jz=0\). None has a root in \(U_j\), so \(V(0,b)=0\) and \(S(0,b)=-q\).
\end{proof}

To determine the frequencies, we first establish a counting identity that also applies to the exponential sums considered in the next section.

\begin{lemma}\label{lem:key-count}
For every fixed \(b\in\F_{q^2}^*\), we have
\begin{equation}\label{eq:key-count}
\#\{a\in\F_{q^2}^*:S(a,b)=2q\}+2\#\{a\in\F_{q^2}^*:S(a,b)=3q\}=\frac{(q+1)^2}{9}.
\end{equation}
\end{lemma}
\begin{proof}
See Appendix A for the complete proof. 
\end{proof}

For a fixed \(c\) with \(1\le c<h\), put \(b=\psi^c\) and use \(t_j=t_j(b)\) from \eqref{eq:shift-tj}. Define
\begin{equation}\label{eq:rho-c}
\rho_c=\#\left\{w\in\F_q^*:\begin{array}{l}
\bigl(R(t_0w),R(t_1w),R(t_2w)\bigr)\text{ is a permutation of }(3,0,0)
\end{array}\right\}.
\end{equation}
Thus \(\rho_c\) is an explicitly specified count over the smaller field \(\F_q\). For each \(w\in\F_q^*\), one determines the numbers of roots in \(\F_q\) of
\[X^3+X=t_jw,\qquad j=0,1,2,\]
and checks whether the three root counts form a permutation of \((3,0,0)\).

The following theorem expresses all frequencies in terms of \(\rho_c\).
\begin{theorem}\label{thm:shift-distribution}
Fix an integer \(c\) with \(1\le c<h\), and let \(\rho_c\) be defined by \eqref{eq:rho-c}. As \(\tau\) ranges from \(0\) to \(q^2-2\), the
cross correlation distribution of \(s\) and \(s'_c\) is
\[
C_\tau(s,s'_c)=
\left\{\begin{array}{lll}
-1,      & \text{occurs }3h^2+2h\rho_c             &\text{times},\\
-1-q,    & \text{occurs }q(q-1)/2-h^2-h\rho_c-1    &\text{times},\\
-1+q,    & \text{occurs }(q+1)(q-2)/6              &\text{times},\\
-1+2q,   & \text{occurs }h^2-2h\rho_c              &\text{times},\\
-1+3q,   & \text{occurs }h\rho_c                   &\text{times}.
\end{array}\right.
\]
\end{theorem}

\begin{proof}
For \(i=-1,0,1,2,3\), define
\[
M_i(c)=\#\{0\le\tau\le q^2-2:C_\tau(s,s'_c)=-1+iq\}=\#\{a\in\F_{q^2}^*:S(a,\psi^c)=iq\}.
\]
Put \(b=\psi^c\). We first show that \(M_3(c)=h\rho_c\). Write \(a=ru\) and \(u^h=\beta^{-\ell}\). The substitution \(y=uz\) transforms the quadratic associated with \(U_j\) into \(y^2+(t_j/r)y+1=0\), with \(y\in U_{j-\ell}\). Lemma~\ref{lem:cubic} therefore shows that \(V(a,b)=4\) if and only if
\[
R(t_\ell/r)=3,\qquad R(t_j/r)=0\quad(j\ne\ell).
\]
As \(r\) runs through \(\F_q^*\), so does \(w=1/r\). The three arguments are then \(t_0w,t_1w,t_2w\). Thus each \(w\) counted by \(\rho_c\) determines exactly one index \(\ell\), and there are exactly \(h\) corresponding choices of \(u\). Conversely, every \(a\) with \(V(a,b)=4\) arises in this way. Hence \(M_3(c)=h\rho_c\).

By Lemma~\ref{lem:key-count},
\[
M_2(c)=h^2-2h\rho_c.
\]
Since \(S(0,b)=-q\), Lemma~\ref{lem:fixed-b-moments} gives
\[
\sum_{i=-1}^{3}M_i(c)=q^2-1,\qquad
\sum_{i=-1}^{3}iM_i(c)=q+1,\qquad
\sum_{i=-1}^{3}i^2M_i(c)=q^2-1.
\]
Substituting the values of \(M_2(c)\) and \(M_3(c)\) determines the other three frequencies as stated.
\end{proof}

Theorem~\ref{thm:shift-distribution} reduces the distribution for each fixed \(c\) to the explicit root count \eqref{eq:rho-c} over \(\F_q\). The formulas are therefore parameterized by \(\rho_c\), whose evaluation for a specified \(c\) gives the numerical frequencies. In particular, \(-1+3q\) occurs if and only if \(\rho_c>0\). We now evaluate the count for two nonzero shifts.

\begin{proposition}\label{prop:rho-normalization}
Suppose \(m=3k\) with \(k\) an odd positive integer, so that \(q=8^k\). Then \(3\mid h\). Fix any root \(\alpha\in\F_8\subseteq\F_q\) of \(X^3+X+1\). For the two shifts \(h/3\) and \(2h/3\) we have
\begin{equation}\label{eq:rho-special-normalization}
 \rho_{h/3}=\rho_{2h/3} =\#\left\{w\in\F_q^*:
 \begin{array}{@{}l@{}}
 \bigl(R(w),R(\alpha w),R((\alpha+1)w)\bigr) \text{ is a permutation of }(3,0,0)
 \end{array}\right\}.
\end{equation}
\end{proposition}

\begin{proof}
Since \(8\equiv-1\pmod 9\) and \(k\) is odd, \(9\mid 8^k+1=q+1\). Thus \(h=(q+1)/3\) is divisible by three. Fix \(c\in\{h/3,2h/3\}\) and put \(b=\psi^c\). From \(\bar b=b^q\) we have \((\bar b/b)^3=\psi^{3c(q-1)}\). Recall that \(\beta=\psi^{(q^2-1)/3}\). Since \(h(q-1)=(q^2-1)/3\), we obtain
\[
 \left(\frac{\bar b}{b}\right)^3=
 \begin{cases}
 \psi^{h(q-1)}=\beta,    & c=h/3,\\  \psi^{2h(q-1)}=\beta^2, & c=2h/3.
 \end{cases}
\]
The two values \(\beta,\beta^2\) are inverses. As \(\beta\) has order three, \(\beta^2+\beta+1=0\), so in either case
\[
 \left(\frac{\bar b}{b}\right)^3+ \left(\frac{b}{\bar b}\right)^3=\beta+\beta^2=1.
\]

Let \(r\) be the unique element of \(\F_q^*\) satisfying \(r^2=b\bar b\). It exists because \(b\bar b=b^{q+1}\in\F_q^*\) and squaring permutes \(\F_q\). In characteristic two,
\[
 (b^3+\bar b^3)^2=b^6+\bar b^6=(b\bar b)^3\left[\left(\frac{b}{\bar b}\right)^3+\left(\frac{\bar b}{b}\right)^3\right]=(b\bar b)^3=(r^3)^2.
\]
Injectivity of squaring gives \(b^3+\bar b^3=r^3\). Now \(t_j=\bar b\beta^j+b\beta^{-j}\) by \eqref{eq:shift-tj}. The product of the two summands is \(b\bar b=r^2\), and their cubes are \(\bar b^3\) and \(b^3\), since \(\beta^3=1\). The identity \((u+v)^3=u^3+v^3+uv(u+v)\) therefore gives
\[
 t_j^3+r^2t_j=r^3,\qquad \left(\frac{t_j}{r}\right)^3+\frac{t_j}{r}=1 \quad(j=0,1,2).
\]
In particular, none of the \(t_j\) is zero. Lemma~\ref{lem:unit-reduction} gives \(t_0+t_1+t_2=0\). If two of them were equal, the third would be zero, so the three roots \(t_0/r,t_1/r,t_2/r\) are distinct.

The cubic \(X^3+X+1\) has no root in \(\F_2\), so it is irreducible over \(\F_2\) and has a root \(\alpha\) in the degree-three extension \(\F_8\). Squaring its equation shows that \(\alpha^2\) is a root, and squaring again gives \(\alpha^4\). Since \(\alpha\ne0,1\) and \(\F_8^*\) has prime order seven, \(\alpha\) has order seven. The exponents \(1,2,4\) are distinct modulo seven, so these three roots exhaust the roots of the cubic. Using \(\alpha^3=\alpha+1\), we get the unordered set equality
\[
 \{t_0,t_1,t_2\}=\{r\alpha,r\alpha^2,r\alpha^4\} =r\alpha\,\{1,\alpha,\alpha+1\}.
\]
In \eqref{eq:rho-c} put \(w'=r\alpha w\). As \(r\alpha\in\F_q^*\), the new variable runs through \(\F_q^*\) exactly once. The three arguments \(t_0w,t_1w,t_2w\) are therefore \(w',\alpha w'\) and \((\alpha+1)w'\) in some order. Reordering them preserves the condition that their root counts form a permutation of \((3,0,0)\). This proves \eqref{eq:rho-special-normalization} for both shifts.
\end{proof}

\begin{theorem}\label{thm:rho-special-shifts}
Suppose \(m=3k\) with \(k\) an odd positive integer. Define the integers \(\kappa_r\) by
\begin{equation}\label{eq:rho-kappa-recurrence}
 \kappa_0=-2,\qquad \kappa_1=3,\qquad \kappa_r=-3\kappa_{r-1}-8\kappa_{r-2}\quad(r\ge2).
\end{equation}
Then \(3\mid h\) and
\begin{equation}\label{eq:rho-special-value}
 \rho_{h/3}=\rho_{2h/3} =\frac{h-\kappa_k}{6} =\frac{q+1-3\kappa_k}{18}.
\end{equation}
For each fixed \(c\in\{h/3,2h/3\}\), the distribution of \(C_\tau(s,s'_c)\) as \(\tau\) ranges from \(0\) to \(q^2-2\) is
\begin{equation}\label{eq:rho-special-frequencies}
C_\tau(s,s'_c)=
\left\{\begin{array}{lll}
-1,      & \text{occurs }(q+1)(10q+10-3\kappa_k)/27             &\text{times},\\
-1-q,    & \text{occurs }(q+1)(20q-61+3\kappa_k)/54    &\text{times},\\
-1+q,    & \text{occurs }(q+1)(q-2)/6              &\text{times},\\
-1+2q,   & \text{occurs }(q+1)(2q+2+3\kappa_k)/27              &\text{times},\\
-1+3q,   & \text{occurs }(q+1)(q+1-3\kappa_k)/54                 &\text{times}.
\end{array}\right.
\end{equation}
\end{theorem}
\begin{proof}
Appendix B first uses invariance under squaring to express \(\rho_c\) in terms of two sums. It then evaluates the triple-root count \(T\) and the mixed sum \(M\) by counting points on two curves over \(\F_8\).
\end{proof}

\begin{example}\label{ex:special-shifts}
For \(m=9\), \(k=3\), \(q=512\) and \(h=171\). By \eqref{eq:rho-kappa-recurrence}, \(\kappa_2=-9+16=7\) and \(\kappa_3=-21-24=-45\). Hence \(\rho_{57}=\rho_{114}=(171+45)/6=36\). For \(m=15\), the next terms are \(\kappa_4=135-56=79\) and \(\kappa_5=-237+360=123\). Here \(h=10923\) and \(\rho_{3641}=\rho_{7282}=(10923-123)/6=1800\).
\end{example}

\begin{remark}
The recurrence \eqref{eq:rho-kappa-recurrence} gives the closed form
\[
 \kappa_r =-\left(\frac{-3+i\sqrt{23}}{2}\right)^r  -\left(\frac{-3-i\sqrt{23}}{2}\right)^r \qquad (r\ge 0).
\]
\end{remark}

\section{Exponential Sums and Related Constacyclic Codes}\label{sec:exponential-codes}
We now determine the complete value distribution of \(S(a,b)\) as \((a,b)\) ranges over \(\F_{q^2}^2\), and use it to construct \(q\)-ary constacyclic codes with explicitly determined weight distributions.
\subsection{Value distribution of binary exponential sum}
\label{subsec:global-S}

The cases in which one coefficient vanishes must be treated separately from the unit-circle bounds for nonzero coefficients.

\begin{lemma}\label{lem:S-exceptional}
We have \(S(0,0)=q^2\), \(S(a,0)=0\) \((a\ne0)\), and, for \(b\ne0\),
\begin{equation}\label{eq:S-axis}
S(0,b)=
\begin{cases}
q(h-1),&b\in D,\\
-q,&b\notin D.
\end{cases}
\end{equation}
Moreover, if \(a,b\ne0\), then
\[
S(a,b)\in\{-q,0,q,2q,3q\}.
\]
\end{lemma}

\begin{proof}
The assertions with \(b=0\) follow from additive-character orthogonality. For \(b\in D\), the other assertions follow from \eqref{eq:S-scaling}, \eqref{eq:S01}, and Lemma~\ref{lem:four-values}. For \(b\in\F_{q^2}^*\setminus D\), write \(b=\psi^c v^d\), where \(1\le c<h\), and apply Lemma~\ref{lem:shift-values} together with \eqref{eq:S-scaling}. 
\end{proof}

The following equation provides the nontrivial solutions needed for the third moment.

\begin{lemma}\label{lem:d-equation}
The solutions in \(\F_{q^2}\) of \(x^d+(x+1)^d=1\) are precisely the elements of \(\F_q\) when \(3\mid m\), and precisely the elements of \(\F_q\cup\{\beta,\beta^2\}\) when \(3\nmid m\). In particular, the equation has \(q-2\varepsilon\) solutions in \(\F_{q^2}\setminus\{0,1\}\), where \(\varepsilon=1\) if \(3\mid m\), and \(\varepsilon=0\) otherwise.
\end{lemma}

\begin{proof}
Every \(x\in\F_q\) is a solution because \(d\equiv1\pmod{q-1}\). Now suppose that \(x\ne0,1\) is a solution, and write
\[
x^d=r_1\beta^j,\qquad (x+1)^d=r_2\beta^k, \qquad r_1,r_2\in\F_q^*,\quad j,k\in\{0,1,2\},
\]
using \(D=\F_q^*\{1,\beta,\beta^2\}\). Since \(\beta\notin\F_q\) and \(\beta^2=\beta+1\), the relation \(r_1\beta^j+r_2\beta^k=1\) has only the following possibilities:
\[
j=k=0,\quad r_1+r_2=1; \qquad\text{or}\qquad \{j,k\}=\{1,2\},\quad r_1=r_2=1.
\]
Indeed, if \(j=k\), the common factor \(\beta^j\) must lie in \(\F_q\). If \(j\ne k\), neither index can be zero, and comparison of the coefficients of \(1,\beta\) gives \(r_1=r_2=1\).

For every \(y\in\F_{q^2}\), the congruence \(d\equiv1\pmod{q-1}\) gives
\begin{equation}\label{eq:norm-power}
(y^d)^{q+1}=(y^{q+1})^d=y^{q+1}.
\end{equation}
In the first case, it follows that
\[
x^{q+1}+(x+1)^{q+1}=r_1^2+r_2^2=1.
\]
The left-hand side is \(x+\bar x+1\), so \(x=\bar x\) and hence \(x\in\F_q\). In the second case, \eqref{eq:norm-power} gives \(x^{q+1}=(x+1)^{q+1}=1\). Thus \(\bar x=x^{-1}\) and \(x+\bar x=1\), whence \(x^2+x+1=0\). Therefore \(x\in\{\beta,\beta^2\}\).

For these last two candidates, the original equation is equivalent to \(\beta^d+\beta^{2d}=1\), which holds if and only if \(3\nmid d\). Since \(d=h\frac{q+2}{2}\), and \(\frac{q+2}{2}\equiv2\pmod3\), we have \(3\mid d\) if and only if \(9\mid q+1\). The order of \(2\) modulo \(9\) is \(6\), so, for odd \(m\), this is equivalent to \(3\mid m\). The result follows.
\end{proof}

\begin{lemma}\label{lem:global-moments}
The first three moments of \(S(a,b)\) are
\begin{equation}\label{eq:global-moments}
\sum_{a,b\in\F_{q^2}}S(a,b)=q^4,\,\,
\sum_{a,b\in\F_{q^2}}S(a,b)^2=q^6,\,\,
\sum_{a,b\in\F_{q^2}}S(a,b)^3=q^4\bigl[3q^2-2+(q^2-1)(q-2\varepsilon)\bigr].
\end{equation}
\end{lemma}

\begin{proof}
The first two identities follow by summing Lemma~\ref{lem:fixed-b-moments} over \(b\). For the third, expand the cube and sum over \(a,b\). Character orthogonality shows that the result is \(q^4\) times the number of triples \((x,y,z)\in\F_{q^2}^3\) satisfying
\[
x+y+z=0,\qquad x^d+y^d+z^d=0.
\]
The triples with at least one zero coordinate consist of \((0,0,0)\) and the three placements of the zero coordinate in a triple with two equal nonzero coordinates. Their number is \(1+3(q^2-1)=3q^2-2\).

If all coordinates are nonzero, put \(t=x/z\). Then \(y=(t+1)z\), where \(t\ne0,1\), and the second equation becomes \(t^d+(t+1)^d=1\). By Lemma~\ref{lem:d-equation}, there are \(q-2\varepsilon\) choices of \(t\), independently of the \(q^2-1\) choices of \(z\). This proves the third identity.
\end{proof}

Combining the preceding moment identities with Lemma~\ref{lem:key-count} gives the complete value distribution of \(S(a,b)\).
\begin{theorem}\label{thm:global-S-distribution}
If \(m\ge3\), the value distribution of \(S(a,b)\), as \((a,b)\) ranges over \(\F_{q^2}^2\), is
\begin{equation}\label{eq:global-S-distribution}
S(a,b)=\left\{\begin{array}{lll}
q^2, &\text{occurs }1 &\text{time},\\
q(h-1), &\text{occurs }3(q-1) &\text{times},\\
-q, & \text{occurs } (q^2-1)\bigl(20q^2+(9\varepsilon-50)q+20\bigr)/54+(q-1)(q-2) &\text{times},\\
0, &\text{occurs } (q^2-1)\bigl(10q^2+(29-9\varepsilon)q-17\bigr)/27 &\text{times},\\
q, & \text{occurs }q(q-2)(q^2-1)/6 &\text{times},\\
2q, & \text{occurs }(q^2-1)\bigl(2q^2+(4+9\varepsilon)q+11\bigr)/27 &\text{times},\\
3q, & \text{occurs }(q^2-1)\bigl(q^2+(2-9\varepsilon)q-8\bigr)/54 &\text{times},
\end{array}
\right.
\end{equation}
where \(\varepsilon=1\) if \(3\mid m\), and \(\varepsilon=0\) otherwise.
\end{theorem}

\begin{proof}
For \(i=-1,0,1,2,3\), let
\[
\nu_i= \#\{(a,b)\in(\F_{q^2}^*)^2:S(a,b)=iq\}.
\]
There are \(3(q-1)\) elements in \(D\) and \((q-1)(q-2)\) in \(\F_{q^2}^*\setminus D\). For the \(k\)-th moment, \(k=1,2,3\), the contribution from pairs with at least one zero coefficient is therefore
\[
q^{2k}+3(q-1)\bigl(q(h-1)\bigr)^k +(q-1)(q-2)(-q)^k.
\]
Subtracting these contributions from the first three moments in Lemma~\ref{lem:global-moments} and dividing the \(k\)-th identity by \(q^k\) gives the three moment equations below. Together with the total number of pairs and the sum of \eqref{eq:key-count} over \(b\ne0\), we obtain
\begin{equation}\label{eq:nonzero-moment-system1-5}
\begin{aligned}
&\nu_2+2\nu_3=(q^2-1)h^2,\\
&\nu_{-1}+\nu_{0}+\nu_{1}+\nu_{2}+\nu_{3}=(q^2-1)^2,\\
&-\nu_{-1}+\nu_{1}+2\nu_{2}+3\nu_{3}=q(q^2-1),\\
&\nu_{-1}+\nu_{1}+4\nu_{2}+9\nu_{3}=(q^2-1)\left(q^2-(q-2)/3\right),\\
&-\nu_{-1}+\nu_{1}+8\nu_{2}+27\nu_{3}=(q^2-1)\left(q(q+2-2\varepsilon)-(q-2)(q-5)/9\right).
\end{aligned}
\end{equation}
Solving this system and adding the contributions in Lemma~\ref{lem:S-exceptional} yields the stated frequencies. For \(m=3\), the two contributions to \(S(a,b)=q(h-1)=2q\) are combined.
\end{proof}

\subsection[A family of q-ary constacyclic codes]{A family of \(q\)-ary constacyclic codes}\label{subsec:constacyclic}

Let \(h_1(X)\) and \(h_2(X)\) be the minimal polynomials of \(\psi\) and \(\psi^d\) over \(\F_q\), respectively. Let \(\widetilde{\mathcal C}\) be the \(q\)-ary cyclic code of length \(q^2-1\) with parity-check polynomial \(h_1(X)h_2(X)\). By the trace representation of cyclic codes,
\begin{equation}\label{eq:parent-cyclic-code}
\widetilde{\mathcal C}=\{\widetilde{\mathbf c}(a,b)=\left(\Tr_m^{2m}\bigl(a\psi^{-i}+b\psi^{-di}\bigr)\right)_{i=0}^{q^2-2}:a,b\in\F_{q^2}\}.
\end{equation}
Retaining the first \(q+1\) coordinates gives the code
\begin{equation}\label{eq:constacyclic-code}
\mathcal C=\{\mathbf c(a,b)=\left(\Tr_m^{2m}\bigl(a\psi^{-i}+b\psi^{-di}\bigr)\right)_{i=0}^{q}:a,b\in\F_{q^2}\}.
\end{equation}

Set \(\lambda=\psi^{q+1}\), which is a primitive element of \(\F_q\). Recall that a linear code of length \(q+1\) over \(\F_q\) is \(\lambda\)-constacyclic if it is invariant under the shift
\[
T_\lambda(c_0,c_1,\ldots,c_q)=(\lambda c_q,c_0,\ldots,c_{q-1}).
\]
We write \(\operatorname{wt}(\mathbf c)\) for the Hamming weight of a codeword \(\mathbf c\).

\begin{theorem}\label{thm:constacyclic-distribution}
The code \(\mathcal C\) is a \(\lambda\)-constacyclic code with parameters \(\left[q+1,\ 4,\ \frac{2(q+1)}3\right]_q.\) Furthermore, its weight distribution is given in Table~\ref{tab:constacyclic-distribution}.
\begin{table}[!ht]
\centering
\caption{The weight distribution of \(\mathcal C\)}
\label{tab:constacyclic-distribution}
\renewcommand{\arraystretch}{1.3}
\begin{tabular}{@{}ll@{}}
\toprule
Weight & Number of codewords\\
\midrule
\(0\)            & \(1\)\\
\(2(q+1)/3\)     & \(3(q-1)\)\\
\(q-3\)          & \((q^2-1)\bigl(q^2+(2-9\varepsilon)q-8\bigr)/54\)\\
\(q-2\)          & \((q^2-1)\bigl(2q^2+(4+9\varepsilon)q+11\bigr)/27\)\\
\(q-1\)          & \(q(q-2)(q^2-1)/6\)\\
\(q\)            & \((q^2-1)\bigl(10q^2+(29-9\varepsilon)q-17\bigr)/27\)\\
\(q+1\)          & \((q^2-1)\bigl(20q^2+(9\varepsilon-50)q+20\bigr)/54+(q-1)(q-2)\)\\
\bottomrule
\end{tabular}
\end{table}
\end{theorem}
\begin{proof}
By Lemma~\ref{lem:exponent}, the multiplicative orders of \(\psi\) and \(\psi^d\) are \(q^2-1\) and \(3(q-1)\), respectively. Neither element belongs to \(\F_q\), so both \(h_1(X)\) and \(h_2(X)\) have degree \(2\). Their different root orders show that these two minimal polynomials are distinct. Therefore,
\[
\dim_{\F_q}\widetilde{\mathcal C}=\deg(h_1h_2)=4.
\]

Since \(d\equiv1\pmod{q-1}\), we have \(\lambda^d=\lambda\). For \(0\le i\le q\) and \(0\le j\le q-2\),
\[
\Tr_m^{2m}\left(a\psi^{-(i+j(q+1))}+b\psi^{-d(i+j(q+1))}\right)=\lambda^{-j}\Tr_m^{2m}\left(a\psi^{-i}+b\psi^{-di}\right).
\]
Consequently,
\begin{equation}\label{eq:cyclic-block-decomposition}
\widetilde{\mathbf c}(a,b)=
\bigl(\mathbf c(a,b),\lambda^{-1}\mathbf c(a,b),
\ldots,
\lambda^{-(q-2)}\mathbf c(a,b)\bigr).
\end{equation}
Thus each word of \(\widetilde{\mathcal C}\) is uniquely determined by its first \(q+1\) coordinates. The coordinate projection from \(\widetilde{\mathcal C}\) onto \(\mathcal C\) is therefore an \(\F_q\)-linear isomorphism, giving \(\dim_{\F_q}\mathcal C=4\). Moreover,
\begin{equation}\label{eq:cyclic-short-weight}
\operatorname{wt}\bigl(\widetilde{\mathbf c}(a,b)\bigr) =(q-1)\operatorname{wt}\bigl(\mathbf c(a,b)\bigr).
\end{equation}

The code \(\mathcal C\) is \(\lambda\)-constacyclic because
\begin{equation}\label{eq:constacyclic-shift}
T_\lambda\bigl(\mathbf c(a,b)\bigr)=\mathbf c(a\psi,b\psi^d).
\end{equation}
Indeed, the first coordinate satisfies
\[
\lambda\Tr_m^{2m}(a\psi^{-q}+b\psi^{-dq})=\Tr_m^{2m}(a\psi+b\psi^d),
\]
where we used \(\psi^{q+1}=(\psi^d)^{q+1}=\lambda\). The remaining coordinates follow by shifting the index.

We next determine the weights through the exponential sum \(S(a,b)\). By additive-character orthogonality over \(\F_q\),
\[
\begin{aligned}
\operatorname{wt}\bigl(\widetilde{\mathbf c}(a,b)\bigr)
&=q^2-\#\left\{x\in\F_{q^2}:\Tr_m^{2m}(ax+bx^d)=0\right\}\\
&=q^2-\frac1q \sum_{y\in\F_q}\sum_{x\in\F_{q^2}}(-1)^{\Tr_1^{2m}(yax+ybx^d)}\\
&=q^2-q-\frac1q \sum_{y\in\F_q^*}\sum_{x\in\F_{q^2}}(-1)^{\Tr_1^{2m}(yax+ybx^d)}.
\end{aligned}
\]
For every \(y\in\F_q^*\), the congruence \(d\equiv1\pmod{q-1}\) gives \(y^d=y\). Hence the substitution \(z=yx\) shows that
\[
\sum_{x\in\F_{q^2}}(-1)^{\Tr_1^{2m}(yax+ybx^d)}=S(a,b).
\]
It follows that
\[
\operatorname{wt}\bigl(\widetilde{\mathbf c}(a,b)\bigr)=(q-1)\left(q-\frac{S(a,b)}q\right).
\]
Combining this identity with \eqref{eq:cyclic-short-weight}, we obtain
\begin{equation}\label{eq:constacyclic-weight}
\operatorname{wt}\bigl(\mathbf c(a,b)\bigr)=q-\frac{S(a,b)}q.
\end{equation}

The parametrization \((a,b)\mapsto\mathbf c(a,b)\) is an \(\F_q\)-linear surjection between two vector spaces of dimension \(4\), and hence is bijective. Therefore, the frequencies in Theorem~\ref{thm:global-S-distribution} translate directly into the numbers of codewords of the corresponding weights in Table~\ref{tab:constacyclic-distribution}.

For odd \(m\ge5\), we have \(q\ge32\), and \(q-3-\frac{2(q+1)}3=\frac{q-11}{3}>0.\) Thus the minimum nonzero weight is \(2(q+1)/3\), and the six nonzero weights are distinct. When \(m=3\), we have \(q=8\) and \(h=3\). The entry of weight \(q-3=5\) has frequency zero, while the two entries of weight \(2h=q-2=6\) have combined frequency \(21+567=588\). In particular, the minimum distance is \(6\), which attains the Singleton bound \(9-4+1=6\).
\end{proof}

\begin{corollary}\label{cor:constacyclic-LCD}
The code \(\mathcal C\) defined in \eqref{eq:constacyclic-code} is Euclidean LCD. For odd \(m\ge5\), it is a six-weight LCD \(\lambda\)-constacyclic code with parameters \(\left[q+1,4,2(q+1)/3\right]_q\). For \(m=3\), it is an \([9,4,6]_8\) LCD MDS \(\lambda\)-constacyclic code.
\end{corollary}
\begin{proof}
We have \(\gcd(q+1,q)=1\). Moreover, \(\lambda=\psi^{q+1}\) has order \(q-1\ge7\), so \(\lambda^2\ne1\). Applying \cite[Corollary~3.3]{LiuFanLiu2018} with \(k=0\) shows that \(\mathcal C\) is Euclidean LCD. The remaining assertions follow from Theorem~\ref{thm:constacyclic-distribution}.
\end{proof}

\section{Conclusion}\label{conclusion}
In this paper, we studied binary sequence cross correlation and related constacyclic codes associated with the Niho exponent \(d=(q-1)(q+4)/6+1\), where \(q=2^m\) and \(m\ge3\) is odd. The main contributions are summarized as follows.
\begin{itemize}
\item
For unshifted decimated sequence \(s'\), we determined the cross correlation distribution explicitly and proved that it has exactly four values: \(-1-q\), \(-1\), \(-1+q\), and \(-1+2q\) (see Theorem~\ref{thm:cross-distribution}).
\item
For each shift decimated sequence \(s'_c\) with \(1\le c<(q+1)/3\), the cross correlation takes at most five values, namely \(-1+iq\) with \(i=-1,0,1,2,3\). We expressed their frequencies in terms of a root-counting parameter \(\rho_c\) over \(\F_q\) (see Theorem~\ref{thm:shift-distribution}).  When \(m=3k\) with \(k\) odd, the two shifts \(c=h/3,2h/3\) both give \(\rho_c=(q+1-3\kappa_k)/18\) (see Theorem~\ref{thm:rho-special-shifts}).
\item
We determined the value distribution of exponential sums \(S(a,b)\) as \((a,b)\) ranges over \(\F_{q^2}^2\) (see Theorem~\ref{thm:global-S-distribution}).
\item
Using relative traces, we constructed \(q\)-ary constacyclic codes with parameters \([q+1,4,2(q+1)/3]_q\) and determined their weight distributions. These codes are Euclidean LCD and have six nonzero weights for \(m\ge5\). For \(m=3\), the construction yields a four-weight \([9,4,6]_8\) LCD MDS code (see Theorem~\ref{thm:constacyclic-distribution} and Corollary~\ref{cor:constacyclic-LCD}).
\end{itemize}

\section*{Appendix A: Proof of Lemma \ref{lem:key-count}}\label{app:key-count}
We prove Lemma \ref{lem:key-count} as follows. We distinguish the cases \(b\in D\) and \(b\notin D\).

\smallskip
\noindent\textit{Case 1: \(b\in D\).}
By the definition of \(D\), there exists \(v\in\F_{q^2}^*\) such that \(v^d=b\). The identity \eqref{eq:S-scaling} gives \(S(a,b)=S(a/v,1).\) Since \(a\mapsto a/v\) permutes \(\F_{q^2}^*\), the distribution for this \(b\), over nonzero \(a\), is the same as that for \(b=1\). By Theorem~\ref{thm:cross-distribution} and the identity \(C_\tau(s,s')=S(\psi^\tau,1)-1\),
\[
\#\{a\in\F_{q^2}^*:S(a,b)=2q\}=\frac{(q+1)^2}{9}=h^2,
\]
whereas \(S(a,b)=3q\) does not occur for \(a\ne0\). This proves \eqref{eq:key-count} in this case.

\smallskip
\noindent\textit{Case 2: \(b\in\F_{q^2}^*\setminus D\).}
Throughout this case, \(b\) is fixed, and we use
\[
t_j=\bar b\beta^j+b\beta^{-j},\qquad j=0,1,2.
\]
By Lemma~\ref{lem:unit-reduction}, these coefficients belong to \(\F_q^*\), are pairwise distinct, and satisfy \(t_0+t_1+t_2=0\). We prove the required identity in five steps.

\smallskip
\noindent\textbf{Step 1: Express the required count in terms of cubic root counts.} For \(r\in\F_q^*\) and \(j\in\{0,1,2\}\), define
\[
A_j(r)=
\begin{cases}
1,&R(t_j/r)=3,\\ 0,&\text{otherwise},
\end{cases}
\qquad
B_j(r)=
\begin{cases}
1,&R(t_j/r)=0,\\ 0,&\text{otherwise}.
\end{cases}
\]
Let \(a\in\F_{q^2}^*\). Its polar representation is \(a=ru\), where \(r\in\F_q^*\) and \(u\in U\). Since \(\bar u=u^{-1}\), we have \(\bar a=ru^{-1}\). There is a unique \(\ell\in\{0,1,2\}\) such that \(u^h=\beta^{-\ell}\).

For each \(j\), the roots contributing to \(V(a,b)\) are the solutions of
\[
az^2+t_jz+\bar a=0,\qquad z\in U_j.
\]
Set \(y=uz\). Then
\[
az^2+t_jz+\bar a=\frac{r}{u}\left(y^2+\frac{t_j}{r}y+1\right), \qquad y^h=u^hz^h=\beta^{j-\ell}.
\]
Thus the contribution from \(U_j\) is the number of roots of
\[
y^2+\frac{t_j}{r}y+1=0
\]
in \(U_{j-\ell}\), where subscripts are read modulo \(3\). Multiplication by \(u\) is a bijection from \(U_j\) to \(U_{j-\ell}\), so this substitution preserves the number of admissible roots.

If \(j=\ell\), the prescribed coset is \(U_0\). By Lemma~\ref{lem:cubic}, it contains both quadratic roots exactly when \(R(t_j/r)=3\), and contains no root otherwise. This contribution is therefore \(2A_\ell(r)\). If \(j\ne\ell\), the prescribed coset is \(U_1\) or \(U_2\). It contains exactly one root when \(R(t_j/r)=0\), and no root otherwise, so the contribution is \(B_j(r)\). Consequently,
\begin{equation}\label{eq:shift-root-count}
V(a,b)=2A_\ell(r)+\sum_{j\ne\ell}B_j(r).
\end{equation}

Recall that \(S(a,b)=q(V(a,b)-1)\). Thus \(S(a,b)=2q\) corresponds to \(V(a,b)=3\), while \(S(a,b)=3q\) corresponds to \(V(a,b)=4\). If \(A_\ell(r)=0\), then \(V(a,b)\le2\), so neither of these values occurs. If \(A_\ell(r)=1\), then the sum \(\sum_{j\ne\ell}B_j(r)\) is \(0\), \(1\), or \(2\), and the corresponding value of \(V(a,b)\) is \(2\), \(3\), or \(4\), respectively. It follows that the contribution to the desired count, with \(V=3\) counted once and \(V=4\) counted twice, is exactly
\[
A_\ell(r)\sum_{j\ne\ell}B_j(r).
\]

For each fixed pair \((r,\ell)\), the equation \(u^h=\beta^{-\ell}\) has \(h\) solutions in \(U\). They give \(h\) distinct elements \(a=ru\), all with the same value in \eqref{eq:shift-root-count}. The uniqueness of the polar representation ensures that different choices of \((r,u)\) give different \(a\). Summing first over \(\ell\) and then over \(r\), we obtain
\begin{equation}\label{eq:key-count-reduced}
\#\{a\in\F_{q^2}^*:S(a,b)=2q\}+2\#\{a\in\F_{q^2}^*:S(a,b)=3q\}=h\sum_{r\in\F_q^*}\sum_{j\ne k}A_j(r)B_k(r),
\end{equation}
where \(j,k\in\{0,1,2\}\) and the inner sum is over the six ordered pairs with \(j\ne k\). It remains to show that the double sum is \(h\).

\smallskip
\noindent\textbf{Step 2: Express the indicators using \(R\) and \(\chi\), and evaluate the basic sums.} By Lemma~\ref{lem:cubic}, the only possibilities for \((R(w),\chi(1/w))\), with \(w\in\F_q^*\), are
\[(1,1),\qquad (3,-1),\qquad (0,-1).\]
Substituting these three pairs shows that \((2R(w)-1-\chi(1/w))/6\) is \(1\) precisely when \(R(w)=3\), and is \(0\) otherwise. Similarly, \((2-R(w)-\chi(1/w))/3\) is \(1\) precisely when \(R(w)=0\). Taking \(w=t_j/r\) gives
\begin{equation}\label{eq:shift-AB}
A_j(r)=\frac{2R(t_j/r)-1-\chi(r/t_j)}6,\quad B_j(r)=\frac{2-R(t_j/r)-\chi(r/t_j)}3.
\end{equation}

Since \(r\mapsto t_j/r\) permutes \(\F_q^*\),
\[
\sum_{r\in\F_q^*}R(t_j/r) =\sum_{w\in\F_q^*}R(w)=q-2.
\]
The last equality follows because \(x^3+x=0\) has exactly the two roots \(0,1\), and every other \(x\in\F_q\) contributes once to a nonzero right-hand side.

For every nonzero \(\alpha\in\F_q\), additive-character orthogonality gives
\[
\sum_{r\in\F_q^*}\chi(\alpha r) =\sum_{r\in\F_q}\chi(\alpha r)-1=-1.
\]
Because \(t_j\ne0\) and \(t_j\ne t_k\) for \(j\ne k\), both \(1/t_j\) and \(1/t_j+1/t_k\) are nonzero. Using also \(\chi(x+y)=\chi(x)\chi(y)\), we conclude that
\begin{equation}\label{eq:shift-basic-sums}
\begin{aligned}
&\sum_{r\in\F_q^*}R(t_j/r)=q-2,\\
&\sum_{r\in\F_q^*}\chi(r/t_j)=-1,\\
&\sum_{r\in\F_q^*}\chi(r/t_j)\chi(r/t_k)=\sum_{r\in\F_q^*}\chi(r/t_j+r/t_k)
=-1\qquad(j\ne k).
\end{aligned}
\end{equation}

\smallskip
\noindent\textbf{Step 3: Evaluate the products of two cubic root counts.} We claim that, for every \(\gamma\in\F_q\setminus\{0,1\}\),
\begin{equation}\label{eq:shift-RR}
\sum_{w\in\F_q^*}R(w)R(\gamma w)=q-4.
\end{equation}
To prove this, count the pairs \((x,y)\in\F_q^2\) satisfying \(y^3+y=\gamma(x^3+x)\). When \(x=0\), we have \(y^3+y=0\), so \(y=0\) or \(1\). This gives two pairs.

For \(x\ne0\), put \(v=y/x\). Substituting \(y=vx\) and dividing by \(x\) gives
\[(v^3+\gamma)x^2=v+\gamma.\]
Since \(m\) is odd, \(\gcd(3,q-1)=1\), so cubing permutes \(\F_q\). There is therefore exactly one \(v\) with \(v^3=\gamma\). This value is different from \(v=\gamma\), since equality would imply \(\gamma^3=\gamma\), or \(\gamma(\gamma+1)^2=0\), contrary to \(\gamma\notin\{0,1\}\).

If \(v^3=\gamma\), the coefficient of \(x^2\) is zero and \(v+\gamma\ne0\), so there is no solution. If \(v=\gamma\), the right-hand side is zero while the coefficient of \(x^2\) is nonzero; the only solution would be \(x=0\), which is excluded. For each of the remaining \(q-2\) choices of \(v\),
\[x^2=\frac{v+\gamma}{v^3+\gamma}\in\F_q^*.\]
Squaring permutes \(\F_q\), so there is exactly one nonzero \(x\), and then \(y=vx\) is uniquely determined. The total number of pairs is therefore \[2+(q-2)=q.\]

Among these pairs, the ones with \(x^3+x=0\) are exactly the four pairs with \(x,y\in\{0,1\}\). For each remaining pair, \(w=x^3+x\) is nonzero, and the number of pairs for a fixed \(w\) is \(R(w)R(\gamma w)\). Removing those four pairs proves \eqref{eq:shift-RR}.

For \(j\ne k\), take \(\gamma=t_k/t_j\). This ratio is neither \(0\) nor \(1\). The substitution \(w=t_j/r\) now gives
\[
\sum_{r\in\F_q^*}R(t_j/r)R(t_k/r)=q-4, \qquad j\ne k.
\]

\smallskip
\noindent\textbf{Step 4: Evaluate the mixed root-count and character terms.} Let \(\gamma\in\F_q\setminus\{0,1\}\). The character identity gives
\[
\chi(\gamma/w)+\chi((\gamma+1)/w) =\chi(\gamma/w)\bigl(1+\chi(1/w)\bigr).
\]
If \(\Tr_1^m(1/w)=1\), the last factor vanishes. If \(\Tr_1^m(1/w)=0\), then \(\chi(1/w)=1\) and \(R(w)=1\), by Lemma~\ref{lem:cubic}. Consequently, after the substitution \(v=1/w\),
\begin{equation}\label{eq:shift-mixed}
\begin{aligned}
&\sum_{w\in\F_q^*}R(w)\bigl(\chi(\gamma/w)+\chi((\gamma+1)/w)\bigr)\\
&\quad=\sum_{w\in\F_q^*}R(w)\chi(\gamma/w)\bigl(1+\chi(1/w)\bigr)\\
&\quad=2\sum_{\substack{v\in\F_q^*, \Tr_1^m(v)=0}}\chi(\gamma v)=-2.
\end{aligned}
\end{equation}
To justify the last equality, the indicator of the condition \(\Tr_1^m(v)=0\) is \((1+\chi(v))/2\). Hence
\[
\begin{aligned}
\sum_{\substack{v\in\F_q, \Tr_1^m(v)=0}}\chi(\gamma v)
&=\frac12\sum_{v\in\F_q}\chi(\gamma v)\bigl(1+\chi(v)\bigr)\\
&=\frac12\left(\sum_{v\in\F_q}\chi(\gamma v)+\sum_{v\in\F_q}\chi((\gamma+1)v)\right)\\
&=0.
\end{aligned}
\]
Both sums vanish because \(\gamma\ne0\) and \(\gamma+1\ne0\). Removing \(v=0\), whose contribution is \(1\), leaves \(-1\). Multiplication by \(2\) proves the value \(-2\) in \eqref{eq:shift-mixed}.

Now fix \(k\in\{0,1,2\}\), and let \(j,\ell\) be the other two indices. Since \(t_0+t_1+t_2=0\),
\[
\frac{t_j}{t_k}+\frac{t_\ell}{t_k}=1.
\] 
Thus these two ratios are \(\gamma\) and \(\gamma+1\), where \(\gamma=t_j/t_k\notin\{0,1\}\). In the sum involving \(R(t_j/r)\), put \(w=t_j/r\); in the sum involving \(R(t_\ell/r)\), put \(w=t_\ell/r\). Both substitutions run through all of \(\F_q^*\). They give
\[
\begin{aligned}
&\sum_{r\in\F_q^*}\bigl(R(t_j/r)+R(t_\ell/r)\bigr)\chi(r/t_k)\\
&\quad=\sum_{w\in\F_q^*}R(w)\left[\chi\left(\frac{t_j}{t_kw}\right)+\chi\left(\frac{t_\ell}{t_kw}\right)\right]\\
&\quad=\sum_{w\in\F_q^*}R(w)\bigl(\chi(\gamma/w)+\chi((\gamma+1)/w)\bigr)\\
&\quad=-2,
\end{aligned}
\]
by \eqref{eq:shift-mixed}. Summing this identity over the three choices of \(k\) includes each ordered pair of distinct indices exactly once. Therefore
\begin{equation}\label{eq:shift-mixed-total}
\sum_{r\in\F_q^*}\sum_{j\ne k} R(t_j/r)\chi(r/t_k)=-6.
\end{equation}

\smallskip
\noindent\textbf{Step 5: Combine the preceding identities.} For fixed \(r\) and \(j\ne k\), equation \eqref{eq:shift-AB} gives
\[
\begin{aligned}
18A_j(r)B_k(r)={}&\bigl(2R(t_j/r)-1-\chi(r/t_j)\bigr)\cdot \bigl(2-R(t_k/r)-\chi(r/t_k)\bigr)\\
={}&4R(t_j/r)-2R(t_j/r)R(t_k/r)-2R(t_j/r)\chi(r/t_k)-2+R(t_k/r)\\
&+\chi(r/t_k)-2\chi(r/t_j)+\chi(r/t_j)R(t_k/r)+\chi(r/t_j)\chi(r/t_k).
\end{aligned}
\]
We sum this expansion over the six ordered pairs \(j\ne k\). Each index appears twice in each position. Thus the terms \(4R(t_j/r)+R(t_k/r)\) give \(10\sum_jR(t_j/r)\), and the terms \(\chi(r/t_k)-2\chi(r/t_j)\) give \(-2\sum_j\chi(r/t_j)\). The constant term \(-2\) gives \(-12\).

For the symmetric products, each unordered pair is counted twice:
\[
\begin{aligned}
\sum_{j\ne k}R(t_j/r)R(t_k/r) &=2\sum_{j<k}R(t_j/r)R(t_k/r),\\
\sum_{j\ne k}\chi(r/t_j)\chi(r/t_k) &=2\sum_{j<k}\chi(r/t_j+r/t_k).
\end{aligned}
\]
For the mixed products, interchanging the two indices gives
\[
\sum_{j\ne k}\chi(r/t_j)R(t_k/r)=\sum_{j\ne k}R(t_j/r)\chi(r/t_k).
\]
The coefficients \(-2\) and \(1\) of these mixed sums therefore combine to \(-1\). We consequently obtain
\[
\begin{aligned}
18\sum_{j\ne k}A_j(r)B_k(r)={}&10\sum_jR(t_j/r)-4\sum_{j<k}R(t_j/r)R(t_k/r)-\sum_{j\ne k}R(t_j/r)\chi(r/t_k)\\
&-12-2\sum_j\chi(r/t_j)+2\sum_{j<k}\chi(r/t_j+r/t_k).
\end{aligned}
\]

Finally, sum over \(r\in\F_q^*\). There are three individual indices and three unordered pairs of distinct indices. Using \eqref{eq:shift-basic-sums}, Step~3, and \eqref{eq:shift-mixed-total}, we obtain
\[
18\sum_{r\in\F_q^*}\sum_{j\ne k}A_j(r)B_k(r)=10\cdot3(q-2)-4\cdot3(q-4)-(-6)-12(q-1)-2\cdot(-3)+2\cdot(-3)=6(q+1).
\]
It follows that
\[
\sum_{r\in\F_q^*}\sum_{j\ne k}A_j(r)B_k(r) =\frac{q+1}{3}=h.
\]
Substituting this into \eqref{eq:key-count-reduced} shows that the required weighted count is \(h\cdot h=h^2\). \hfill\(\square\)

\section*{Appendix B: Proof of Theorem~\ref{thm:rho-special-shifts}}
\label{app:rho-special-proof}

\noindent\textbf{Step 1. Express \(\rho_c\) in terms of two sums.} Write \(q=8^k\) with \(k\) odd, fix \(c\in\{h/3,2h/3\}\), and choose \(\alpha\) as in Proposition~\ref{prop:rho-normalization}. By \eqref{eq:rho-special-normalization}, we may take the three right-hand sides in the order \(w,\alpha w,(\alpha+1)w\). We will use squaring to show that the three products of two root counts and one character have the same sum over \(w\). For this computation and those below, \(\alpha^3=\alpha+1\) gives
\[
\begin{array}{c|cccc}
j      &     3   &  4  &  5  &   6\\ \hline
\alpha^j  &\alpha+1  &\alpha^2+\alpha  &\alpha^2+\alpha+1  &  \alpha^2+1
\end{array}
\]

Fix \(w\in\F_q^*\). At each of these three right-hand sides we record the number of distinct roots \(x\in\F_q\) and the character of the reciprocal of the right-hand side. In the same order these are
\[
\begin{aligned}
\bigl(R_0(w),R_1(w),R_2(w)\bigr)  &=\bigl(R(w),R(\alpha w),R((\alpha+1)w)\bigr),\\
\bigl(C_0(w),C_1(w),C_2(w)\bigr)  &=\bigl(\chi(1/w),\chi(1/(\alpha w)),\chi(1/((\alpha+1)w))\bigr).
\end{aligned}
\]
For example, \(R_1(w)=\#\{x\in\F_q:x^3+x=\alpha w\}\). The subscript labels one of the three equations for the same \(w\). For both shifts, \(\rho_c\) counts precisely the \(w\) for which \((R_0(w),R_1(w),R_2(w))\) is one of \((3,0,0),(0,3,0),(0,0,3)\).

Squaring gives a bijection from the roots for \(a\) to the roots for \(a^2\), and the absolute trace is invariant under squaring. Hence \(R(a^2)=R(a)\) and \(\chi(1/a^2)=\chi(1/a)\) for \(a\ne0\). With \(w'=w^2/\alpha\), the squared arguments are
\[
(w^2,\alpha^2w^2,(\alpha+1)^2w^2)=\bigl(\alpha w',(\alpha+1)w',w'\bigr).
\]
Here \(\alpha^3=\alpha+1\) and \(\alpha(\alpha+1)^2=\alpha^7=1\). As \(w\mapsto w'\) permutes \(\F_q^*\), cyclically permuting the three positions preserves the sum of any product of the corresponding root counts and characters. In particular, the three products obtained by placing one \(C_i\) and two \(R_i\)'s in distinct positions have the same sum.

In the following expansion we suppress the argument \(w\), writing \(R_i=R_i(w)\) and \(C_i=C_i(w)\) inside each summand. Define
\begin{equation}\label{eq:rho-T-M-definitions}
T=\sum_{w\in\F_q^*}R_0R_1R_2,\qquad M=\sum_{w\in\F_q^*}R_0R_1C_2 .
\end{equation}

Lemma~\ref{lem:cubic} and \eqref{eq:shift-AB} give the following indicators, where the two right-hand columns are evaluated over the rationals.
\[
\begin{array}{cc|cc}
R_i &C_i &\mathbf{1}_{R_i=3}=(2R_i-1-C_i)/6& \mathbf{1}_{R_i=0}=(2-R_i-C_i)/3\\ \hline
1   & 1  &  0 & 0\\
3   &-1  &  1 & 0\\
0   &-1  &  0 & 1
\end{array}
\]
If position \(i\) contains the \(3\), let \(j,\ell\) be the other two positions. Its contribution has denominator \(6\cdot3\cdot3=54\) and numerator \((2R_i-1-C_i)(2-R_j-C_j)(2-R_\ell-C_\ell)\). The three possible positions are disjoint. Summing their numerators and collecting terms gives
\[
\begin{aligned}
54\rho_c=\sum_{w\ne0}\Bigl[
&6R_0R_1R_2+3(C_0R_1R_2+C_1R_0R_2+C_2R_0R_1)+12\sum_iR_i\\
&-9\sum_{i<j}R_iR_j-3\sum_{i\ne j}R_iC_j-12+3\sum_{i<j}C_iC_j-3C_0C_1C_2\Bigr].
\end{aligned}
\]
For example, the three products contribute \(-1,2,2\) to the coefficient of \(C_0R_1R_2\), and \(2,-1,-1\) to that of \(R_0C_1C_2\). This gives the coefficient \(3\) in the first case and cancels the term in the second.

Each single root count sums to \(q-2\). Indeed, its right-hand side runs over \(\F_q^*\), so the sum counts each \(x\) once except for the two zeros \(0,1\) of \(x^3+x\). Each unordered pair of root counts sums to \(q-4\) by \eqref{eq:shift-RR}. Each of the three products with one character and two root counts sums to \(M\) by the cyclic symmetry above. For the six ordered pairs of distinct indices, \eqref{eq:shift-mixed-total} gives \(\sum_{w\ne0}\sum_{i\ne j}R_iC_j=-6\). It applies here because \(1,\alpha,\alpha+1\) are nonzero, distinct, and sum to zero. The change \(r=1/w\) matches their arguments with those in Appendix A.

For \(i\ne j\), the product \(C_iC_j\) has the form \(\chi(\delta/w)\) with \(\delta\ne0\), since the two reciprocal coefficients are distinct. For \(C_0C_1C_2\), the coefficient is \(1+\alpha^{-1}+(\alpha+1)^{-1}=1+\alpha^6+\alpha^4=\alpha\ne0\). Each of these products therefore sums to \(-1\) by additive-character orthogonality. The total of the terms other than \(6T+9M\) in \(54\rho_c\) is
\[
 36(q-2)-27(q-4)-3(-6)-12(q-1)-9+3=60-3q .
\]
Division by three gives
\begin{equation}\label{eq:rho-indicator-reduction}
18\rho_c=2T+3M-q+20 .
\end{equation}

\medskip
\noindent\textbf{Step 2. Evaluate \(T\).}
Put \(f(X)=X^3+X\). For a fixed \(w\ne0\), the three factors in \(T\) count independent choices of \(x,y,z\) with \(f(x)=w,\ f(y)=\alpha w,\ f(z)=(\alpha+1)w\). The first coordinate recovers \(w\). Thus \(T\) counts exactly the ordered triples \((x,y,z)\in\F_q^3\) satisfying
\begin{equation}\label{eq:ordered}
 f(y)=\alpha f(x),\qquad f(z)=(\alpha+1)f(x),\qquad f(x)\ne0 .
\end{equation}
Since \(f(x)=x(x+1)^2\ne0\), we have \(x\ne0,1\). Set \(v=y/x\) and \(u=z/x\). Substitution into the two equations and division by \(x\) give
\[
 (v^3+\alpha)x^2=v+\alpha,\qquad (u^3+\alpha+1)x^2=u+\alpha+1 .
\]
Write \(\theta=\alpha^5\), so \(\theta^3=\alpha\). Cubing permutes \(\F_q\) because \(3\nmid q-1\). At \(v=\theta\) the first equation has zero left side and nonzero right side. At \(v=\alpha\) it forces \(x=0\), and at \(v=0,1\) it forces \(x=1\). For every other \(v\) the first equation gives
\begin{equation}\label{eq:slope}
 x^2=\frac{v+\alpha}{v^3+\alpha},\qquad x^2+1=\frac{v(v+1)^2}{v^3+\alpha},\qquad y=vx .
\end{equation}
Both fractions are nonzero. As squaring is bijective, this recovers a unique \(x\ne0,1\) and the corresponding \(y\).

Eliminating \(x^2\) from the two equations gives \((v+\alpha)(u^3+\alpha+1)+(u+\alpha+1)(v^3+\alpha)=0\), or
\begin{equation}\label{eq:rho-special-quartic}
 vu^3+v^3u+\alpha(u^3+u)+(\alpha+1)(v^3+v)=0 .
\end{equation}
For an admissible \(v\), division by \(v^3+\alpha\) recovers the second equation for \(x^2\) without any division by \(u^3+\alpha+1\). Hence each such affine point \((v,u)\) gives exactly one triple, with \(z=ux\). Substitution of the four excluded \(v\)'s gives
\[
\begin{array}{c|l|l}
v&\text{equation in }u&\text{possible }u\\ \hline
0      & \alpha(u^3+u)=0                  &0,1\\
1      & (\alpha+1)(u^3+u)=0              &0,1\\
\alpha & (\alpha^3+\alpha)(u+\alpha+1)=0  &\alpha+1\\
\theta & (\theta+\alpha)(u^3+\alpha+1)=0  &\alpha
\end{array}
\]
Here \(\alpha^3+\alpha=1\), and \(\alpha\) is the unique cube root of \(\alpha+1\). Therefore \eqref{eq:rho-special-quartic} has \(T+2+2+1+1=T+6\) affine points over \(\F_q\).

For \(v\ne1\), put \(t=(u+1)/(v+1)\) and \(\eta=v+1\). Substituting \(v=1+\eta,\ u=1+t\eta\) into \eqref{eq:rho-special-quartic} and dividing by \(\eta^2\) gives, with \(Q(t)=(t+\theta)(t^2+t+1)\),
\begin{equation}\label{eq:rho-special-quadratic}
 t(t+1)^2\eta^2+(\alpha+1)Q(t)\eta+(\alpha+1)(t^2+\theta)=0 .
\end{equation}
This change removes the two points \((v,u)=(1,0),(1,1)\). The new equation therefore has \(T+4\) solutions with \(\eta\ne0\). At \(\eta=0\) it requires \(t^2=\theta\), whose unique solution is \(t=\alpha^6\). Thus its total affine count is \(T+5\).

For \(t(t+1)Q(t)\ne0\), we first scale \(\eta\) so that its quadratic and linear terms take the form \(Y_0^2+Y_0\). Multiplying \eqref{eq:rho-special-quadratic} by \(t(t+1)^2/((\alpha+1)^2Q(t)^2)\) gives
\[
 Y_0=\frac{t(t+1)^2\eta}{(\alpha+1)Q(t)},\qquad Y_0^2+Y_0=\frac{t(t+1)^2(t^2+\theta)}{(\alpha+1)Q(t)^2}.
\]
To cancel one factor of \(Q(t)\) in the denominator, we seek a square with the same remainder modulo \(Q(t)\) as the numerator \(t(t+1)^2(t^2+\theta)/(\alpha+1)\). Polynomial division gives the two remainders
\[
\begin{aligned}
\frac{t(t+1)^2(t^2+\theta)}{\alpha+1}
   &\equiv \theta t^2+\alpha^6t+\alpha^3 \pmod{Q(t)},\\
t^4&\equiv \alpha^2t^2+\alpha^6t+\alpha^2 \pmod{Q(t)}.
\end{aligned}
\]
Adding cancels the linear term and leaves \((\alpha+1)t^2+\theta\). Thus the numerator is congruent to \(t^4+(\alpha+1)t^2+\theta\), which equals \((t^2+\theta t+\alpha^6)^2\) because \(\theta^2=\alpha+1\) and \((\alpha^6)^2=\theta\). Keeping the quotient as well gives the identity
\[
\frac{t(t+1)^2(t^2+\theta)}{\alpha+1}  =(t^2+\theta t+\alpha^6)^2+Q(t)\bigl(\alpha^4t^2+(\alpha+1)t+1\bigr).
\]
In characteristic two, replacing \(Y_0\) by \(Y=Y_0+s\) adds \(s^2+s\) to the right-hand side. Thus the square term above is cancelled by taking
\[
Y=\frac{t(t+1)^2\eta}{(\alpha+1)Q(t)}+\frac{t^2+\theta t+\alpha^6}{Q(t)} .
\]
The remaining numerator over \(Q(t)\) is \(\alpha^4t^2+(\alpha+1)t+1+(t^2+\theta t+\alpha^6)\). Collecting its coefficients gives
\begin{equation}\label{eq:first-curve}
 Y^2+Y=\frac{\theta t^2+\alpha^2t+\alpha^2}{Q(t)} .
\end{equation}
For each such \(t\) the coefficient of \(\eta\) in the change is nonzero, so it gives a bijection of the two sets of solutions.

Let \(\mathcal X\) be the smooth projective model of \eqref{eq:first-curve} over \(\F_8\). At \(t=\theta\), the numerator on the right-hand side equals \(\theta^3+\alpha^2\theta+\alpha^2=\alpha+1+\alpha^2=\theta\ne0\). Modulo \(t^2+t+1\), we have \(t^2\equiv t+1\), so the numerator reduces to \((\theta+\alpha^2)(t+1)=(\alpha+1)(t+1)\). The two roots of \(t^2+t+1\) are distinct, since its derivative is \(1\). Both have order three, since \((t+1)(t^2+t+1)=t^3+1\) and \(1\) is not a root of the quadratic. As \(\F_8^*\) has order seven, both roots lie outside \(\F_8\). Thus neither is \(\theta\), and the numerator is nonzero at both roots. The right-hand side therefore has three simple poles over the algebraic closure. Its value at infinity is zero because the numerator has degree two and the denominator has degree three.

If a rational function \(H\) has a pole of order \(s>0\), then \(H^2\) and \(H\) have pole orders \(2s\) and \(s\) there, so \(H^2+H\) has pole order \(2s\). The simple poles of the right-hand side of \eqref{eq:first-curve} therefore rule out a root in \(\overline{\F_8}(t)\). The quadratic in \(Y\) is irreducible over this field and has derivative \(1\). Hence \(\mathcal X\) is geometrically irreducible, and projection onto the \(t\)-line gives a separable cover of degree two.

We apply \cite[Prop.~3.7.8(b)--(d)]{Stichtenoth2009} to this Artin--Schreier cover. A simple pole has order one, which is prime to the characteristic, so the integer \(m_P\) in that proposition equals \(1\) there. At a regular place the same definition gives \(m_P=-1\). Parts (b) and (c) show that each pole is totally ramified and that all regular places are unramified. Above a rational simple pole there is a unique point, with ramification index two. Its ramification index times its residue degree equals the degree of the cover, namely two. Its residue degree is therefore one and the point is rational over the same field.

For the genus calculation in part (d), the base is the rational function field \(\F_8(t)\), of genus zero. A finite place defined by an irreducible polynomial \(P(t)\) has residue field \(\F_8[t]/(P(t))\), of degree \(\deg P\) over \(\F_8\). Its contribution to the sum in part (d) is \((m_P+1)\deg P\). Every regular place thus contributes zero. At the poles, the factor \(t+\theta\) has degree one and the irreducible factor \(t^2+t+1\) has degree two. Substitution into part (d) gives
\[
g(\mathcal X)   =2\cdot0+\frac{2-1}{2}  \bigl(-2+(1+1)\cdot1+(1+1)\cdot2\bigr)=\frac{-2+2+4}{2}=2.
\]
Equivalently, if the right-hand side has exactly \(n\) simple poles over the algebraic closure and no other poles, all \(n\) geometric poles have degree one and the same formula gives \(g=(-2+2n)/2=n-1\).

Since \(k\) is odd, \(q=8^k\equiv-1\pmod3\). Thus \(\F_q^*\) contains no element of order three, so \(t^2+t+1\) has no root in \(\F_q\). The two counts can differ only at the four rational parameters in the table.
\[
\begin{array}{c|cccc}
 t&0&1&\theta&\infty\\ \hline
 \text{affine solutions of \eqref{eq:rho-special-quadratic}}&1&1&1&0\\
 \text{points on }\mathcal X&2&2&1&2
\end{array}
\]
For the first row, \(t=0,1\) each give \(\eta=1\), and at \(t=\theta\) the equation determines \(\eta^2\) uniquely. For the second row, the right side of \eqref{eq:first-curve} takes the values \(\alpha^4,\alpha,0\) at \(0,1,\infty\), respectively. They all have absolute trace zero already in \(\F_8\), since \(\alpha+\alpha^2+\alpha^4=0\). Thus \(Y^2+Y\) takes each of these values twice in \(\F_8\) and also in \(\F_q\). At the simple pole \(\theta\) there is one point. It follows that
\begin{equation}\label{eq:rho-special-T-point-count}
 \#\mathcal X(\F_q)=(T+5)+(2-1)+(2-1)+(1-1)+(2-0)=T+9 .
\end{equation}

Over \(\F_8\), we can count all points directly using \eqref{eq:trace-fibres}. The trace of \(a_0+a_1\alpha+a_2\alpha^2\), with \(a_i\in\F_2\), is \(a_0\), since \(1\) has trace one and \(\alpha,\alpha^2\) have trace zero. This gives
\[
\begin{array}{c|ccccccccc}
 t&0&1&\alpha&\alpha^2&\alpha^3&\alpha^4&\alpha^5&\alpha^6&\infty\\ \hline
 \text{right side of \eqref{eq:first-curve}}
     &\alpha^4&\alpha&1&\alpha^5&\alpha^6&\alpha^3&\mathrm{pole}&\alpha^2&0\\
 \text{points}&2&2&0&0&0&0&1&2&2
\end{array}
\]
The last row sums to \(9\).

We recall how to pass to the odd extensions. For a smooth projective geometrically irreducible curve of genus \(g\) over \(\F_8\), write \(N_r\) for its number of \(\F_{8^r}\)-points. Its zeta function satisfies \cite[Sec.~5.1, Thm.~5.1.15 and Cor.~5.1.16]{Stichtenoth2009}
\[
\exp\!\left(\sum_{r\ge1}N_r\,\frac{Z^r}{r}\right) =\frac{L(Z)}{(1-Z)(1-8Z)},\qquad  L(Z)=\sum_{j=0}^{2g}c_jZ^j,\quad c_0=1.
\]
The functional equation \(L(Z)=8^g Z^{2g}L(1/(8Z))\) gives
\begin{equation}\label{eq:rho-L-functional}
c_{2g-j}=8^{g-j}c_j\qquad(0\le j\le g).
\end{equation}
Taking the logarithmic derivative and expanding \(1/(1-Z)\) and \(8/(1-8Z)\) as geometric series gives the identity of formal power series over \(\mathbb Q\) used below and in Step~4 \cite[Cor.~5.1.17(a)]{Stichtenoth2009},
\begin{equation}\label{eq:rho-L-point-count}
 \frac{L'(Z)}{L(Z)} =\sum_{r\ge1}\bigl(N_r-(8^r+1)\bigr)Z^{r-1}.
\end{equation}
For \(\mathcal X\), the constant term in this identity is \(c_1=9-9=0\). As \(g=2\), \eqref{eq:rho-L-functional} gives \(c_3=8c_1=0\) and \(c_4=64\). Hence \(L(Z)=1+c_2Z^2+64Z^4\) is even. Its derivative has only odd powers, and \(1/L(Z)\) has only even powers. Thus the coefficient of \(Z^{k-1}\) in \(L'(Z)/L(Z)\) vanishes for every odd \(k\). Equation~\eqref{eq:rho-L-point-count} gives \(\#\mathcal X(\F_{8^k})=8^k+1=q+1\). Comparing with \eqref{eq:rho-special-T-point-count}, we obtain
\begin{equation}\label{eq:rho-special-T}
T=q+1-9=q-8.
\end{equation}

\medskip
\noindent\textbf{Step 3. Reduce \(M\) to an additive character sum.}
For a fixed \(w\ne0\), the product \(R_0(w)R_1(w)\) counts the ordered pairs with \(f(x)=w,\ f(y)=\alpha w\). Summing over \(w\in\F_q^*\) counts each pair with \(f(x)\ne0\) and \(f(y)=\alpha f(x)\) exactly once, since \(w=f(x)\) is then determined. On this whole set of pairs, \eqref{eq:slope} makes \(v=y/x\) a bijection onto \(\F_q\setminus\{0,1,\alpha,\theta\}\). Its inverse recovers \(x\) from \(x^2\) and sets \(y=vx\). It remains to express the weight \(C_2(w)=\chi(1/((\alpha+1)f(x)))\) in terms of this \(v\).

Since \(x\ne0,1\) and \((\alpha+1)^{-1}=\alpha^4\), use
\[
\frac{1}{f(x)}=\frac{1}{x(x+1)^2} =\frac1x+\frac1{x+1}+\frac1{(x+1)^2}.
\]
Inside the absolute trace, \(\alpha^4/(x+1)^2\) can be replaced by \(\alpha^2/(x+1)\), since the first is the square of the second. Now \(\alpha^4+\alpha^2=\alpha\). Squaring the resulting argument once more preserves its trace, so
\[
\chi\!\left(\frac{1}{(\alpha+1)f(x)}\right)
=\chi\!\left(\frac{\alpha^4}{x}+\frac{\alpha}{x+1}\right)
=\chi\!\left(\frac{\alpha}{x^2}+\frac{\alpha^2}{x^2+1}\right).
\]
After substituting \eqref{eq:slope}, the last argument becomes
\begin{equation}\label{eq:before-trace}
 \alpha\,\frac{v^3+\alpha}{v+\alpha}+\alpha^2\,\frac{v^3+\alpha}{v(v+1)^2}.
\end{equation}
The necessary partial fractions are
\[
\begin{aligned}
 \frac{v^3+\alpha}{v+\alpha}   &=v^2+\alpha v+\alpha^2+\frac1{v+\alpha},\\
 \frac{v^3+\alpha}{v(v+1)^2}   &=1+\frac{\alpha}{v}+\frac{\alpha}{v+1}+\frac{\alpha+1}{(v+1)^2}.
\end{aligned}
\]
In the first line the remainder is \(\alpha^3+\alpha=1\). In the second, the three proper fractions have combined numerator \(\alpha(v+1)^2+\alpha v(v+1)+(\alpha+1)v=v+\alpha\). In \eqref{eq:before-trace} the square terms \(\alpha v^2\) and \(\alpha^5/(v+1)^2\) have the same traces as \(\alpha^4v\) and \(\alpha^6/(v+1)\), respectively. Collecting terms, the coefficient of \(v\) is \(\alpha^4+\alpha^2=\alpha\), the constant is \(\alpha^3+\alpha^2=\theta\), and the coefficient of \(1/(v+1)\) is \(\alpha^3+\alpha^6=\alpha^4\). Thus each admissible slope contributes \(\chi(F(v))\), where
\begin{equation}\label{eq:mixed-function}
 F(v)=\alpha v+\theta+\frac{\alpha^3}{v} +\frac{\alpha^4}{v+1}+\frac{\alpha}{v+\alpha}.
\end{equation}

Of the four excluded slopes, \(0,1,\alpha\) are poles of \(F\). The remaining slope \(\theta\) is regular. Since \(\theta+1=\alpha^4\) and \(\theta+\alpha=\alpha^6\), we obtain
\[
 F(\theta)=\alpha^6+\theta+\alpha^5+1+\alpha^2=\alpha^6+1+\alpha^2=0 .
\]
Its character is \(+1\). Putting this single term back gives the sum over all finite regular values,
\begin{equation}\label{eq:rho-special-character-sum}
M+1=\sum_{v\in\F_q\setminus\{0,1,\alpha\}}\chi(F(v)).
\end{equation}

\medskip
\noindent\textbf{Step 4. Evaluate the character sum.}
Let \(\mathcal Y\) be the smooth projective model of \(Y^2+Y=F(v)\) over \(\F_8\), and now write \(N_r=\#\mathcal Y(\F_{8^r})\). The three fractions in \eqref{eq:mixed-function} have simple poles at \(0,1,\alpha\), respectively, and the term \(\alpha v\) gives a simple pole at infinity. These four poles are distinct and rational.  As in Step~2, a simple pole rules out \(F=H^2+H\) for every \(H\in\overline{\F_8}(v)\). Thus \(\mathcal Y\) is geometrically irreducible. By the simple-pole formula used in Step~2, \(\mathcal Y\) has genus three and one rational point over each pole. At each of the \(8^r-3\) finite regular parameters, \eqref{eq:trace-fibres} gives \(1+(-1)^{\Tr_1^{3r}(F(v))}\) points. Adding the four pole points yields
\begin{equation}\label{eq:rho-genus-three-point-count}
 N_r=8^r+1+\sum_{v\in\F_{8^r}\setminus\{0,1,\alpha\}} (-1)^{\Tr_1^{3r}(F(v))}.
\end{equation}
At \(r=k\), \eqref{eq:rho-special-character-sum} gives \(N_k=q+1+(M+1)=q+2+M\). Substituting this and \(T=q-8\) into \eqref{eq:rho-indicator-reduction} gives
\begin{equation}\label{eq:rho-special-proof-identity}
 18\rho_c=q+4+3M=q+1+3\bigl(N_k-(q+1)\bigr).
\end{equation}

The numerator \(L\) for \(\mathcal Y\) has degree six. By \eqref{eq:rho-L-functional}, \(c_4=8c_2,\ c_5=64c_1,\ c_6=512\). It therefore suffices to determine \(c_1,c_2,c_3\). For this we need \(N_1,N_2,N_3\). The degree \(r=2\) is used here to recover a coefficient of \(L\). The restriction that \(k\) be odd applies when using \eqref{eq:rho-special-proof-identity}.

By \eqref{eq:trace-fibres}, a regular parameter \(v\) gives two points if \(\Tr_1^{3r}(F(v))=0\), and none otherwise. Hence
\[
 N_r=4+2\,\#\{v\in\F_{8^r}\setminus\{0,1,\alpha\}: \Tr_1^{3r}(F(v))=0\}.
\]
The initial \(4\) accounts for the three finite poles and the pole at infinity.

We compute these three finite counts in the polynomial bases shown below. In each row \(e=3r\), \(P\) is irreducible of degree \(e\) over \(\F_2\), and \(z\) denotes the residue of the indeterminate in \(\F_2[z]/(P)\). The displayed \(\alpha\) satisfies \(\alpha^3+\alpha+1=0\). Different choices of its conjugate give the same point count, since squaring both coordinates gives a bijection with the curve having squared coefficients. For each \(v\notin\{0,1,\alpha\}\), reduce \(F(v)\) modulo \(P\) and write \(F(v)=\sum_{j=0}^{e-1}u_jz^j\) with \(u_j\in\F_2\). The fourth column gives its trace directly from these coefficients. All additions in that column are in \(\F_2\).
\[
\begin{array}{c|c|c|c|r|r|r}
 r&P&\alpha&\Tr_1^e(F(v))&\#\,\text{trace }0&\#\,\text{trace }1&N_r\\ \hline
 1&z^3+z+1&z&u_0&1&4&6\\
 2&z^6+z+1&z^3+z^2+z&u_5&41&20&86\\
 3&z^9+z^4+1&z^4+z^3+z^2&u_0+u_5&277&232&558
\end{array}
\]

The first trace rule was proved in Step~2. To verify the other two, use the \(\F_2\)-linearity of the trace. The traces of the basis elements are determined at once by the formal expansion at infinity
\[
\frac{P'(X)}{P(X)}  =\sum_{i=0}^{e-1}\frac{1}{X-z^{\,2^i}} =\sum_{j\ge0}\Tr_1^e(z^j)\,X^{-j-1}.
\]
The first equality follows by factoring \(P\) into its distinct linear factors. The second follows by expanding each fraction as a geometric series in \(X^{-1}\). For \(e=6\), the quotient \(1/(X^6+X+1)\) starts with \(X^{-6}\), so among \(1,z,\ldots,z^5\) only \(z^5\) has trace one. For \(e=9\), the quotient is \(X^{-1}/(1+X^{-5}+X^{-9})\). Its only nonzero terms through \(X^{-9}\) are \(X^{-1}\) and \(X^{-6}\), so precisely \(1\) and \(z^5\) have trace one. This proves the two remaining trace rules.

For example, in the first row the five regular values give
\[
\begin{array}{c|ccccc}
 v&\alpha^2&\alpha^3&\alpha^4&\alpha^5&\alpha^6\\ \hline
 F(v)&\alpha^5&1&\alpha^6&0&\alpha^5\\
 \Tr_1^3(F(v))&1&1&1&0&1
\end{array}
\]
Indeed, at \(v=\alpha^3\) the five terms of \eqref{eq:mixed-function} are \(\alpha^4,\alpha^5,1,\alpha^3,\alpha\), whose sum is \(1\). Thus only \(v=\alpha^5\) contributes two points, giving \(N_1=4+2=6\). In the larger fields the same substitution uses division modulo \(P\) by nonzero elements. For instance, at \(v=z\) in the second row one obtains \(F(z)=z^3+z\), whose \(z^5\)-coefficient is zero. The two trace columns account for all \(5,61,509\) finite regular parameters in their respective rows. Their zero-trace counts give \(N_2=4+2\cdot41=86\) and \(N_3=4+2\cdot277=558\).

The first three differences \(N_r-(8^r+1)\) are \(6-9=-3\), \(86-65=21\) and \(558-513=45\). Multiply \eqref{eq:rho-L-point-count} by \(L(Z)\) and compare the coefficients of \(1,Z,Z^2\). This gives
\[
\begin{aligned}
 c_1&=-3,\\
 2c_2&=21-3c_1=21+9=30, & c_2&=15,\\
 3c_3&=45+21c_1-3c_2=45-63-45=-63, & c_3&=-21 .
\end{aligned}
\]
Completing the coefficients by \eqref{eq:rho-L-functional} and factoring over the integers gives
\begin{equation}\label{eq:rho-genus-three-L}
\begin{aligned}
 L(Z)&=1-3Z+15Z^2-21Z^3+120Z^4-192Z^5+512Z^6\\
     &=(1-3Z+8Z^2)^2(1+3Z+8Z^2).
\end{aligned}
\end{equation}

Let \(\omega_+,\omega_-\) be the roots of \(X^2+3X+8\). Their sum is \(-3\) and their product is \(8\), so
\[
 (1-\omega_+Z)(1-\omega_-Z)=1+3Z+8Z^2,\qquad
 (1+\omega_+Z)(1+\omega_-Z)=1-3Z+8Z^2 .
\]
By \cite[Cor.~5.1.16]{Stichtenoth2009}, \(N_r-(8^r+1)\) is minus the
sum of the \(r\)-th powers of the reciprocal roots of \(L\), counted with
multiplicity. Here a reciprocal root \(\omega\) corresponds to a factor
\(1-\omega Z\). To see the sign and the multiplicities directly from
\eqref{eq:rho-L-point-count}, one copy of this factor contributes
\[
 \frac{(1-\omega Z)'}{1-\omega Z}
   =-\frac{\omega}{1-\omega Z}
   =-\sum_{r\ge1}\omega^rZ^{r-1}
\]
to \(L'/L\). A repeated factor contributes this series once for each
copy. In \eqref{eq:rho-genus-three-L}, the factor \(1+3Z+8Z^2\) gives
\(\omega_+,\omega_-\) once each, and \((1-3Z+8Z^2)^2\) gives
\(-\omega_+,-\omega_-\) twice each. Comparing the coefficients of
\(Z^{r-1}\) therefore gives
\[
\begin{aligned}
 N_r-(8^r+1)
   &=-\bigl(\omega_+^r+\omega_-^r
           +2(-\omega_+)^r+2(-\omega_-)^r\bigr)\\
   &=-\bigl(1+2(-1)^r\bigr)(\omega_+^r+\omega_-^r).
\end{aligned}
\]
The numbers \(-(\omega_+^r+\omega_-^r)\) start at \(-2,3\) for \(r=0,1\). Multiplying \(\omega_\pm^2=-3\omega_\pm-8\) by \(\omega_\pm^{r-2}\) shows that these numbers satisfy the recurrence \(\kappa_r=-3\kappa_{r-1}-8\kappa_{r-2}\) in \eqref{eq:rho-kappa-recurrence}. Hence \(N_r-(8^r+1)=(1+2(-1)^r)\kappa_r\). For our odd \(k\), the factor \(1+2(-1)^k\) is \(-1\). It follows that \(N_k=q+1-\kappa_k\) and \(M=N_k-q-2=-1-\kappa_k\). Substitution into \eqref{eq:rho-special-proof-identity} yields
\[
 \rho_{h/3}=\rho_{2h/3} =\frac{q+1-3\kappa_k}{18}=\frac{h-\kappa_k}{6}.
\]
Finally substitute \(\rho_c=(h-\kappa_k)/6\) and \(q=3h-1\) into Theorem~\ref{thm:shift-distribution}.

\end{document}